\documentclass[fleqn,usenatbib]{mnras}

\usepackage{graphicx}
\usepackage{amsmath}
\usepackage{amssymb}
\usepackage[T1]{fontenc}
\usepackage{txfonts}

\newcommand{\etafit}{\ensuremath{\eta_\mathrm{fit}}}
\newcommand{\ebprp}{\ensuremath{E(G_\mathrm{BP}{-}G_\mathrm{RP})}}
\newcommand{\logage}{\ensuremath{\log(\mathrm{Age/yr})}}
\newcommand{\feh}{\ensuremath{[\mathrm{Fe/H}]}}

\newcommand{\dynesty}{\textsc{dynesty}}

\graphicspath{{figures/}{./}}

\title[Bayesian NS Parameters for 5056 Open Clusters]{Astrophysical Parameters
       of 5056 Open Star Clusters from Bayesian Nested Sampling with PARSEC
       Isochrones}

\author[O. Plevne \& F. Akbaba]{
Olcay Plevne,$^{1,2}$\thanks{E-mail: olcayplevne@istanbul.edu.tr}\thanks{T\"{U}B\.{I}TAK 2219 International Postdoctoral Research Fellow.}
and Furkan Akbaba$^{3}$
\\
$^{1}$Faculty of Science, Department of Astronomy and Space Sciences,
Istanbul University, Istanbul, Turkey\\
$^{2}$Institute for Astronomy, University of Edinburgh, Royal Observatory,
Blackford Hill, Edinburgh, EH9 3HJ, UK\\
$^{3}$Institute of Graduate Studies in Science, Istanbul University,
Istanbul, Turkey
}

\date{Accepted XXXX. Received YYYY; in original form ZZZZ}
\pubyear{2026}

\usepackage{hyperref}   

\begin{document}
\label{firstpage}
\pagerange{\pageref{firstpage}--\pageref{lastpage}}
\maketitle

\begin{abstract}
We present a homogeneous catalogue of ages, metallicities, distances, and
reddenings for $5\,056$ open clusters in the Unified Cluster Catalogue (UCC),
derived uniformly from \textit{Gaia} Data Release 3 (DR3) colour--magnitude
diagrams via Bayesian nested sampling with PARSEC isochrones. Unlike previous
large-scale catalogues, metallicity is fitted as a free parameter for every
cluster rather than fixed or adopted from the literature: where available,
\textit{Gaia} XP spectrophotometric metallicities inform the fit ($1\,201$
clusters), while the remaining clusters are constrained by broad priors from
\textit{Gaia} parallaxes and the Schlegel--Finkbeiner--Davis dust map. No
cluster is anchored to an external open-cluster parameter catalogue, and no
per-cluster manual intervention is applied. A fit-quality metric, \etafit,
identifies $3\,807$ clusters ($75$~per~cent) with reliable colour--magnitude
solutions, spanning ages of $0.004$--$5.9$\,Gyr and distances out to
$17$\,kpc. We validate the resulting parameters against six literature
catalogues and, more critically, against independent high-resolution
spectroscopic metallicities from OCCAM DR19: for $132$ clusters in common we
find a median absolute deviation of only $0.054$\,dex ($-0.021 \pm 0.132$\,dex
mean offset), comparable to the intrinsic systematic uncertainty of
isochrone-based metallicities. As a direct consequence of this fidelity, the
radial metallicity gradient recovered from our XP-informed clusters
($-0.064$\,dex\,kpc$^{-1}$) steepens towards the spectroscopically calibrated
value ($-0.079$\,dex\,kpc$^{-1}$) for the highest-quality subset -- demonstrating
that a purely photometric pipeline, built without reference to spectroscopy,
recovers a physically meaningful abundance scale. The full catalogue and
posterior chains are publicly released.
\end{abstract}



\section{Introduction}
\label{sec:intro}

Open star clusters are among the most powerful tracers of Galactic structure and chemical
evolution. Born within the same molecular cloud, their member stars share a common age,
distance, metallicity, and initial chemical composition, making them ideal standard candles
and chronometers for mapping the Milky Way disc. Accurate determinations of their
astrophysical parameters --- age, metallicity, distance, and interstellar reddening ---
underpin a broad range of investigations including the age--metallicity relation of the
thin disc \citep{Tarricq2021}, the chemical abundance structure of the disc
\citep{Otto2026}, the star-formation history of the solar neighbourhood, and the
calibration of stellar evolution models.

The systematic cataloguing of open clusters has a long history. The catalogue of
\citet{Dias2002} (DAML), continuously updated for two decades, served as the
primary community reference with more than 2\,000 entries. \citet{Kharchenko2013}
extended this effort with the Milky Way Star Clusters catalogue (MWSC), providing
homogeneous parameters for 3\,784 clusters from 2MASS near-infrared photometry using
isochrone fitting. Despite their value, both compilations suffered from heterogeneous input
data and non-uniform fitting procedures, which introduced systematic offsets between
sub-samples.

The advent of \textit{Gaia} transformed open cluster science. The precise parallaxes,
proper motions, and broad-band photometry delivered by \textit{Gaia} Data Release 2
\citep{GaiaCollab2018} and Data Release 3 \citep{GaiaCollab2023} enabled the separation
of cluster members from field stars with unprecedented fidelity. An early
\textit{Gaia} DR2 effort was the Bayesian isochrone analysis of
\citet{Bossini2019}, who derived homogeneous ages, distances, and extinctions
for 269 well-populated clusters with the BASE-9 code. \citet{CantatGaudin2020} used \textit{Gaia}
DR2 data to derive membership lists and astrophysical parameters for 1\,867 clusters via
an artificial neural network trained on PARSEC isochrones at fixed solar metallicity,
so that [Fe/H] was not a free parameter.
\citet{Dias2021} subsequently fitted isochrones to 1\,743 clusters using a
Bayesian approach that includes metallicity as a free parameter, but adopts a
position-dependent prior based on an assumed Galactic abundance--radius relation; consequently
the recovered [Fe/H] values are partly anchored to that assumed relation
rather than being driven solely by each cluster's CMD.
The combination of HDBSCAN clustering and \textit{Gaia} DR3 photometry allowed
\citet{Hunt2021} and \citet{Hunt2023} to push the open-cluster
census to 7\,167 objects. \citet{Hunt2023} derive ages, distances, and
extinctions with a neural-network fitter operating within a fixed metallicity range,
but do not report photometric [Fe/H] from the isochrone fit.
\citet{Hunt2024} further refined this sample using cluster masses and
dynamics. More recently, \citet{Cavallo2024} applied an artificial neural
network to \textit{Gaia}~DR3 colour--magnitude diagrams to derive ages,
distances, extinctions, and photometric metallicities for $\sim$5\,400 open
clusters, one of the few \textit{Gaia}-era catalogues to provide a free
photometric [Fe/H] at large scale, albeit as point estimates without full
posterior distributions. A comprehensive review of the open cluster population
in the \textit{Gaia} era is given by \citet{CantatGaudin2022}.

Most recently, \citet{Perren2023} introduced Unified Cluster Catalogue (UCC)%
\footnote{\url{https://ucc.ar}}, aggregating entries from numerous independent studies into a single homogeneous database of more than 18\,000 stellar clusters. The UCC provides 5D phase-space membership lists derived with the FASTMP \citep{Perren2023} algorithm and assigns a C3 quality flag with sixteen classes ranging from AA (most reliable) to DD, encoding the level of independent confirmation of the membership solution. With this large, well-characterised membership base, the UCC provides an unprecedented opportunity to derive astrophysical parameters for thousands of clusters using a single, self-consistent methodology.

Bayesian inference offers several advantages over classical grid-search or
$\chi^2$-minimisation isochrone fitting. By sampling the full posterior distribution of
model parameters, it naturally propagates photometric uncertainties and membership
probabilities into credible intervals on derived quantities, avoids degeneracies by
including physically motivated priors, and provides a quantitative model-fit quality
metric \citep{Skilling2004,Speagle2020}. The PARSEC stellar evolution
models \citep{Bressan2012,Marigo2017} are widely used for isochrone
fitting owing to their coverage of the Hertzsprung--Russell diagram from the pre-main
sequence through the asymptotic giant branch and their publicly accessible online interface.

In this paper we present a homogeneous catalogue of astrophysical parameters for
$5\,056$ open clusters selected from the AA and AB quality classes of the UCC.
These two classes represent the highest-confidence membership solutions in the UCC
and were chosen specifically because well-defined membership lists are a prerequisite
for reliable isochrone fitting; the resulting sample size is a consequence of this
quality-first selection rather than a target in itself. The parameters are derived by
Bayesian Nested Sampling (NS) with PARSEC isochrones fitted to \textit{Gaia} DR3
colour--magnitude diagrams (CMDs). A central motivation is to address a gap in
existing large catalogues: \citet{CantatGaudin2020} fix metallicity to the solar
value; \citet{Dias2021} include metallicity but adopt a
position-dependent abundance--radius relation as the metallicity prior, so the resulting
[Fe/H] values are not fully independent of that assumed relation; \citet{Hunt2023} do not report photometric [Fe/H] from the isochrone
fit at all; and \citet{Cavallo2024}, while providing photometric [Fe/H] for $\sim$5\,400 clusters via an artificial neural network trained on synthetic CMDs, do not deliver formal posterior probability distributions for the fitted parameters. Our pipeline treats initial metallicity $Z_\mathrm{ini}$ as a fully free
parameter across the entire grid, yielding a photometric [Fe/H] posterior for every
cluster that is not anchored to an assumed abundance model. Where \textit{Gaia} XP
spectrophotometric metallicities are available ($N = 1\,201$ clusters in our sample), we fold them in
as an \emph{informative starting point} for the prior rather than a hard constraint:
the nested sampler explores the full prior volume and the XP information merely
narrows the initial search region, so the likelihood can move the posterior away
from the prior mean if the CMD demands it. It is for these clusters that the present
catalogue is most powerful: we deliver a genuinely informative photometric \feh{}
posterior that combines the \textit{Gaia} XP measurement with the full CMD evidence
in a single self-consistent inference. \textit{Gaia}'s XP spectrophotometry is thus
the key enabling resource of this work --- it promotes cluster metallicity from an
assumed input to a measured quantity for a substantial, and with \textit{Gaia} DR4
rapidly expanding, fraction of the census. For the remaining clusters a broad,
environment-dependent uniform prior is adopted; there the optical CMD alone provides
only limited metallicity discrimination, and the resulting [Fe/H] posteriors should be
interpreted as broad, CMD-modulated constraints rather than precise individual
measurements (quantified in Section~\ref{sec:results:properties}). The
degree of CMD constraint for every cluster is reported through the inlier fraction $\eta_\mathrm{fit}$,
which flags clusters where the isochrone fit did not converge to a well-defined
sequence and the reported parameters should be treated with appropriate caution. For the
high-quality subset, the derived parameters span ages from $\sim$4\,Myr to $\sim$5.9\,Gyr, distances from 86 to
17\,012\,pc, and metallicities from $-1.17$ to $+0.46$\,dex.

This paper is organised as follows. Section~\ref{sec:data} describes the input data,
including the UCC membership catalogue and the auxiliary datasets used to construct
informative priors. Section~\ref{sec:method} presents the Bayesian NS methodology, the
PARSEC isochrone grid, and the likelihood function. Section~\ref{sec:results} describes
the resulting parameter catalogue and its statistical properties. Section~\ref{sec:comparison}
compares our results with selected literature catalogues.
Section~\ref{sec:conclusions} summarises the main conclusions.

\begin{figure*}
  \centering
  \includegraphics[width=\textwidth]{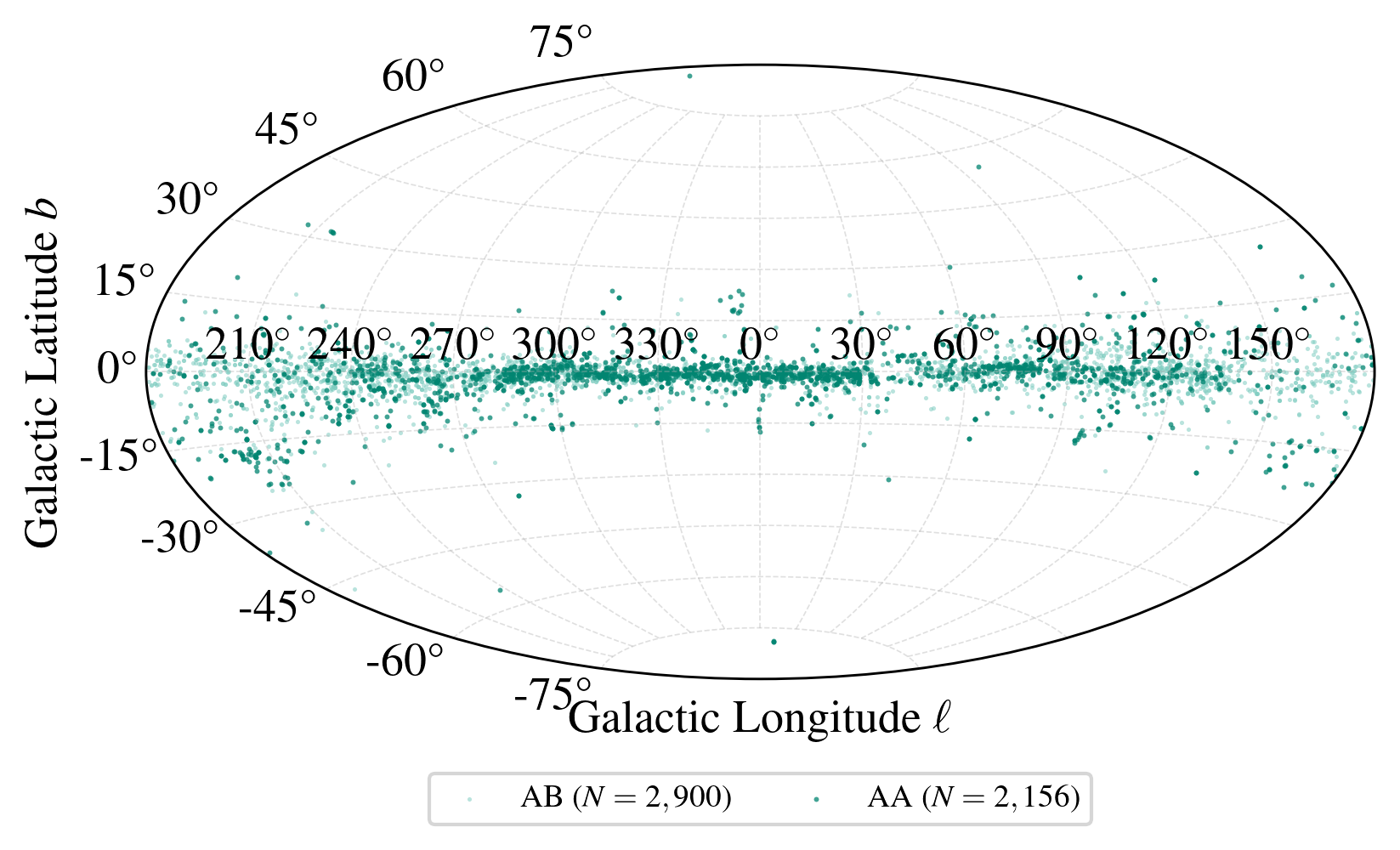}
   \caption{Aitoff projection of the sky distribution in Galactic coordinates ($l, b$) of the 5056 open clusters processed in this work. The Galactic centre is at the projection centre; longitude increases to the left. Class AA clusters ($N = 2156$, dark teal) represent the highest tier where both the photometric score ($C_{\mathrm{phot}}$) and density score ($C_{\mathrm{dens}}$) are in the top segment. Class AB clusters ($N = 2900$, light teal) possess the highest photometric quality but a slightly lower density score ($C_{\mathrm{dens}} \in [0.50, 0.75]$), following the UCC C3 quality classification framework \citep{Perren2023}. The strong concentration towards the Galactic plane ($|b| \le 10^{\circ}$) reflects the thin-disc origin of open clusters, with the sample extending across all longitudes to trace both the solar neighbourhood and the outer disc.}
  \label{fig:spatial}
\end{figure*}

\section{Data}
\label{sec:data}

\subsection{The Unified Cluster Catalogue and \textit{Gaia} DR3 membership}
\label{sec:data:ucc}

The primary input to our analysis is the Unified Cluster Catalogue
\citep[UCC;][]{Perren2023}, which consolidates more than 18\,000 Galactic stellar clusters
from dozens of independent discovery and membership studies into a single homogeneous
database. For each cluster the UCC provides celestial and Galactic coordinates, median
parallax, proper motions, and --- crucially --- a list of probable member stars with
individual membership probabilities derived by the FASTMP algorithm \citep{Perren2023} operating in the
5-dimensional \textit{Gaia} phase space ($l$, $b$, $\varpi$, $\mu_{\alpha*}$, $\mu_\delta$).
The \textit{Gaia} DR3 photometry \citep{GaiaCollab2023} provides the $G$, $G_\mathrm{BP}$,
and $G_\mathrm{RP}$ magnitudes used to construct the colour--magnitude diagrams fitted in
this work.

Clusters in the UCC are assigned one of sixteen quality classes (C3), ranging from AA to DD \citep{Perren2023}.
The C3 class combines two independent quality metrics: a photometry-based score ($C_\mathrm{phot}$) and a density-based score ($C_\mathrm{dens}$) each binned into four segments labelled A (highest, $[0.75, 1]$) through D (lowest, $[0, 0.25]$); the first letter encodes ($C_\mathrm{phot}$) and the second ($C_\mathrm{dens}$) \citep{Perren2023}. Class AA clusters therefore have both metrics in the highest tier; class AB clusters have the highest photometric quality but a slightly lower density score ($C_\mathrm{dens} \in [0.50, 0.75]$).
Together these two classes represent the highest-reliability tier of the UCC.
Because accurate isochrone fitting requires well-defined, high-purity membership
lists, we deliberately restrict our analysis to these classes; the final sample
size of 5\,056 clusters is a direct outcome of this quality-first selection criterion.
The combined AA and AB classes of the UCC contain $5\,591$ clusters.
We select those that retain at least five probable member stars after
applying the membership probability threshold described below; $535$
clusters are excluded by this criterion, leaving a working sample of
$5\,056$ clusters ($2\,156$ class AA and $2\,900$ class AB;
Fig.~\ref{fig:spatial}). Our member-star catalogue contains approximately one million
star entries across all processed clusters.

For each cluster we retain only stars with membership probability
$\mathrm{probs} \geq 0.75$, providing a high-purity photometric sample while preserving
sufficient stars for isochrone fitting even in low-richness systems. This threshold is
consistent with practices adopted in recent \textit{Gaia}-era studies
\citep{Dias2021,Hunt2023}.

\begin{figure*}
  \centering
  \includegraphics[width=\textwidth]{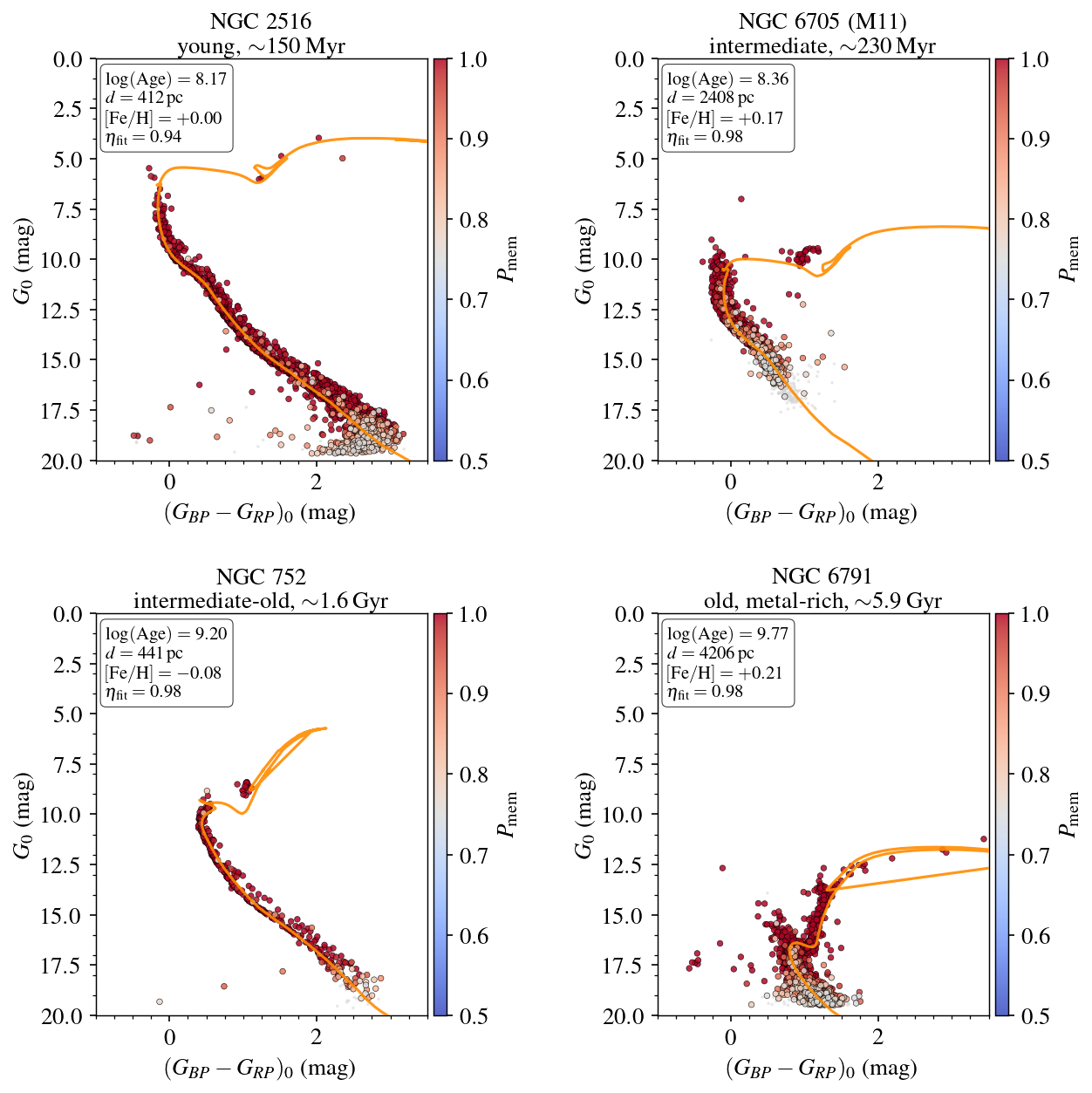}
  \caption{Dereddened colour--magnitude diagrams for four representative
           clusters spanning the full age range of the catalogue.
           Stars are colour-coded by membership probability ($P_\mathrm{mem}$,
           coolwarm scale, $0.5$--$1.0$) with black outlines; grey points are
           stars below the $p \geq 0.75$ threshold.
           The orange curve shows the best-fitting PARSEC isochrone at the
           posterior median parameters.
           Key parameters are annotated in each panel:
           $\log(\mathrm{Age/yr})$, heliocentric distance $d$,
           photometric metallicity \feh, and fit quality $\etafit$.
           The $y$-axis shows the apparent dereddened magnitude $G_0 = G - A_G$;
           the $x$-axis shows the dereddened colour $(G_\mathrm{BP}-G_\mathrm{RP})_0$.
           \textit{Top left:} NGC\,2516 (young, $\sim$150\,Myr);
           \textit{top right:} NGC\,6705 = M11 (intermediate, $\sim$230\,Myr);
           \textit{bottom left:} NGC\,752 (intermediate-old, $\sim$1.6\,Gyr);
           \textit{bottom right:} NGC\,6791 (old, metal-rich, $\sim$5.9\,Gyr).}
  \label{fig:example_cmds}
\end{figure*}

\subsection{Auxiliary datasets and prior information}
\label{sec:data:aux}

\subsubsection*{Distance priors from \textit{Gaia} parallaxes}

For each cluster we compute a distance prior centred on the median parallax-inverted
distance of its member stars, propagating the uncertainty as a truncated Gaussian
with a lower bound at zero distance.
For every cluster in our working sample the distance prior is derived from the
median of the Gaia DR3 parallax-inverted distances of its member stars; parallax
is thus the basis of the distance prior throughout the catalogue.

\subsubsection*{Metallicity priors from \textit{Gaia} XP spectra}

\textit{Gaia} DR3 \citep{GaiaCollab2023} provides low-resolution XP spectra from
which stellar metallicities can be estimated. We adopt the data-driven
metallicities of \citet{Andrae2023}, who train an XGBoost model on \textit{Gaia}
XP spectra combined with CatWISE infrared photometry to derive the overall
metallicity $[\mathrm{M/H}]$ for $\sim$175 million stars; following their recommendation we use only member
giants ($\log g < 3.5$), for which the XP metallicities are most reliable. For
clusters with at least three such member stars
having XP-based $[\mathrm{Fe/H}]$ measurements we compute the median cluster metallicity
and its dispersion, and adopt a truncated Gaussian as the $[\mathrm{Fe/H}]$ prior.
Strictly, \citet{Andrae2023} provide $[\mathrm{M/H}]$ rather than
$[\mathrm{Fe/H}]$; for the predominantly thin-disc, near-solar-metallicity open
clusters studied here, $[\mathrm{M/H}] \approx [\mathrm{Fe/H}]$ is a reasonable
approximation, and we adopt it throughout when referring to the XP-based prior. This
criterion is met by 1\,346 clusters with \textit{Gaia} XP metallicities; $1\,201$ of
these fall within our final AA/AB sample (644 AA and 557 AB) and receive the XP Gaussian
prior, of which $798$ also satisfy the high-quality fit criterion $\eta_\mathrm{fit} \geq 0.80$
(see Section~\ref{sec:results:properties}). It is important to emphasise that
this XP-based Gaussian serves only as an \emph{informative starting point} that narrows the
initial search region; it does not act as a hard constraint. The nested sampler explores
the full prior volume, and the likelihood can --- and does --- displace the posterior away
from the XP prior mean whenever the CMD evidence demands it. For the remaining clusters we
adopt a uniform prior over $[\mathrm{Fe/H}]$ whose bounds depend on the cluster's
Galactocentric position: $[-0.5, +0.5]$\,dex for the solar-neighbourhood and standard disc \citep{Hayden2015, Donor2020, Netopil2022},
$[-0.3, +0.5]$\,dex for the inner Galactic disc ($R_\mathrm{GC} < 5$\,kpc) \citep{Spina2021}, and
$[-1.5, 0.0]$\,dex for the outer disc ($d > 15$\,kpc or $|b| > 30\degr$) \citep{Otto2026}.

\subsubsection*{Reddening priors from the SFD dust map}

Line-of-sight $E(B{-}V)$ values for each cluster are drawn from the
\citet{Schlegel1998} Schlegel--Finkbeiner--Davis (SFD) dust map. These are
converted to $E(G_\mathrm{BP}{-}G_\mathrm{RP})$ using the coefficient $K_{E(G_\mathrm{BP}{-}G_\mathrm{RP})/E(B{-}V)} = 1.305$
\citep{Wang2019} and serve as the mean of a truncated Gaussian reddening prior,
corrected for the finite distance of each cluster so as not to over-estimate the
foreground dust column. The SFD value thus provides a physically motivated upper envelope
for the reddening prior rather than a rigid constraint.

\subsubsection*{Age prior}

Age priors are uniform in $\log(\mathrm{Age/yr})$ over the range $[6.0, 10.0]$ for all
clusters. A broad, uninformative prior is deliberately adopted to avoid conditioning the
posterior on heterogeneous literature age estimates, which carry their own systematic
uncertainties.

\begin{figure}
  \centering
  \includegraphics[width=\columnwidth]{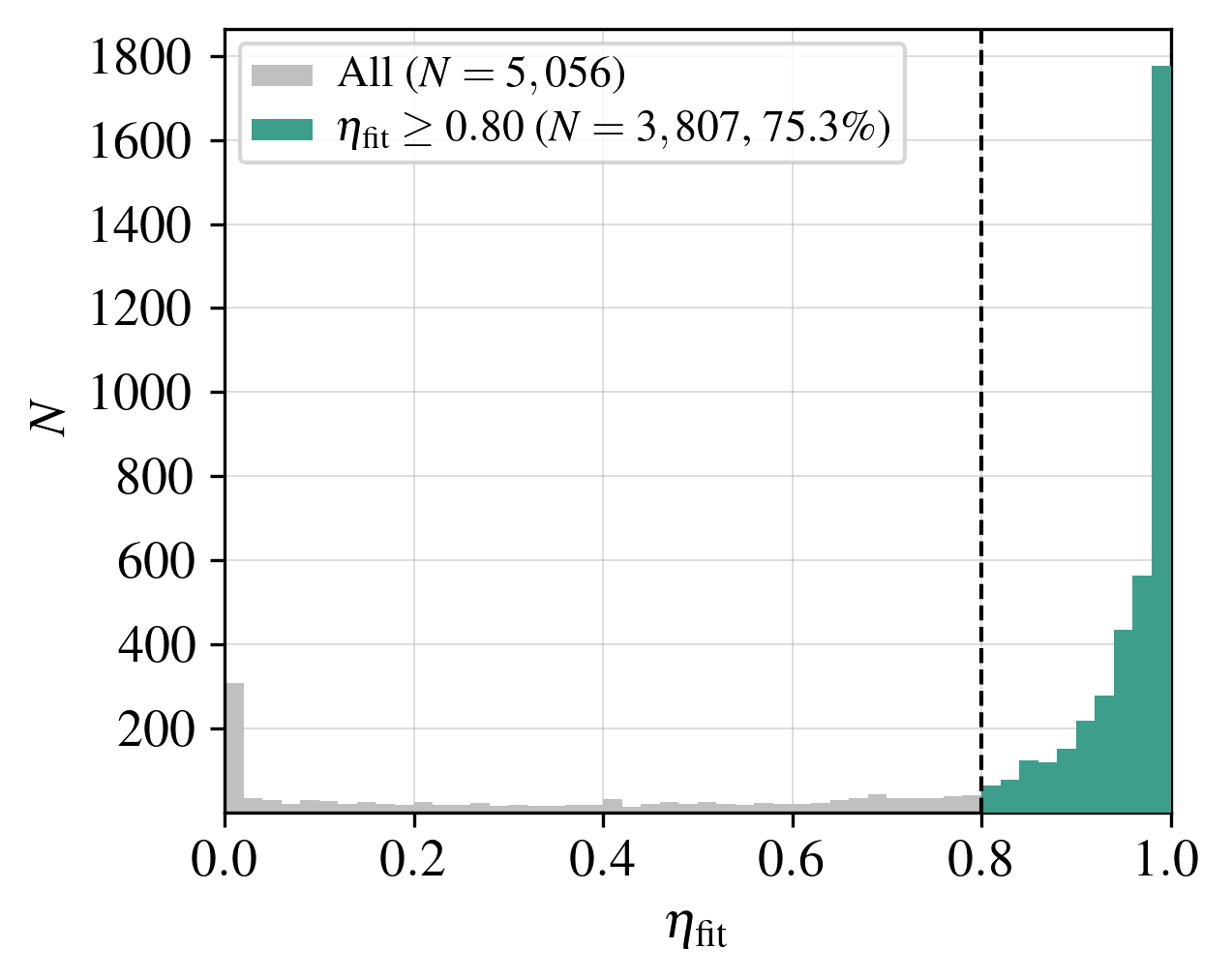}
  \caption{Distribution of the fit quality indicator $\etafit$ for all
           $5\,056$ clusters (grey) and the high-quality subset
           ($\etafit \geq 0.80$, teal; $N = 3\,807$, $75.3$\,per\,cent).
           The dashed vertical line marks the adopted quality threshold.
           The distribution peaks sharply near $\etafit \approx 0.95$,
           indicating that the majority of the sample is well-constrained,
           while the tail below $0.80$ captures clusters with degenerate or
           poorly populated CMDs.}
  \label{fig:etafit_hist}
\end{figure}

\section{Method}
\label{sec:method}

We determine the astrophysical parameters of each open cluster through a Bayesian inference framework that compares extinction-corrected colour--magnitude diagrams (CMDs) of probable member stars with theoretical stellar isochrones. The pipeline is applied uniformly across the full sample, with no manual intervention on individual clusters.

The simultaneous, posterior-based approach adopted here was first developed and
validated on individual open clusters by \citet{TanikOzturk2025} and
\citet{Karagoz2025}, who combined ground-based CCD $UBV$ photometry with
\textit{Gaia} DR3 astrometry and used a Markov chain Monte Carlo (MCMC) sampler
to estimate cluster reddening, metallicity, distance, and age \emph{jointly},
rather than sequentially as in classical isochrone fitting. More recently,
\citet{Yontan2026} applied the same MCMC-based joint-fitting approach, using
\textit{Gaia} DR3 photometry and astrometry alone, to a homogeneous sample of
27 King open clusters. Extending this
methodology from a handful of individually analysed clusters to a catalogue of
thousands motivated two changes. First, the fit is based on \textit{Gaia} DR3
CMDs alone (Section~\ref{sec:data}), since homogeneous multi-band ground-based
photometry is not available for the full sample. Second, and more importantly,
the MCMC sampler is replaced by nested sampling \citep{Skilling2004, Speagle2020}.
For an unsupervised pipeline applied to $5\,056$ clusters, nested sampling offers
decisive practical advantages: it explores the multimodal and strongly
degenerate posteriors typical of sparse or heavily reddened CMDs without
per-cluster tuning of proposal scales, burn-in, or convergence diagnostics; it
returns the Bayesian evidence $\ln Z$ as a by-product; and its termination is set
by a single objective stopping criterion applied identically to every cluster.
This transition from individual-cluster MCMC studies to a fully automated,
posterior-sampling catalogue is what makes the present analysis both tractable
and reproducible at scale. A nested-sampling fitting scheme of this kind
was first tested in our analysis of the multiple populations of
$\omega$~Centauri \citep{Akbaba2026omegaCen}.

\subsection{Isochrone grid and interpolation}
\label{sec:isochrones}

We use PARSEC version 1.2S stellar isochrones \citep{Bressan2012, Chen2014, Marigo2017} computed in the \textit{Gaia} EDR3 photometric system \citep{GaiaCollab2021}, covering the $G$, $G_\mathrm{BP}$, and $G_\mathrm{RP}$ passbands. The grid spans $\log(\mathrm{Age/yr}) = 6.0$--$10.0$, with a step of $0.05$\,dex for younger populations ($\log(\mathrm{Age/yr}) \leq 8.5$) and $0.10$\,dex for older populations, and covers $[\mathrm{Fe/H}] = -1.5$ to $+0.4$\,dex in steps of $0.025$\,dex (corresponding to $Z_\mathrm{ini} \approx 0.0005$--$0.038$ at $Z_\odot = 0.0152$), yielding a total of $5\,050$ isochrones that comprise $\sim$$622{,}000$ individual CMD points ($\sim$$123$ points per isochrone along the main sequence, subgiant, and red-giant branches). At the start of each run the entire grid is loaded into RAM in compressed \textsc{parquet} format. For a given set of model parameters $(\log\mathrm{Age}, Z_\mathrm{ini})$, the nearest grid point is located with a \textsc{cKDTree} query. CMD coordinates at arbitrary initial stellar mass are then computed with a \textsc{DecisionTree} interpolator trained on the grid, which maps $(Z_\mathrm{ini}, \log\mathrm{Age}, M_\mathrm{ini})$ to the corresponding $(G_\mathrm{BP}-G_\mathrm{RP}, G)$ values. This two-stage approach (nearest-grid selection followed by in-node interpolation) keeps isochrone access time below a millisecond per likelihood evaluation while preserving smooth CMD morphology.

\subsection{Bayesian model and priors}
\label{sec:priors}

The model has four free parameters: logarithmic age $\theta_1 = \log(\mathrm{Age/yr})$, initial metallicity $\theta_2 = Z_\mathrm{ini}$, heliocentric distance $\theta_3 = d$ (pc), and reddening $\theta_4 = E(G_{BP}-G_{RP})$.

\subsubsection*{Age}
A broad uniform prior,
\begin{equation}
    p(\log\mathrm{Age}) = \mathcal{U}(6.0,\,10.0),
\end{equation}
is adopted for all clusters. We deliberately do not condition the age prior on literature catalogues in order to avoid propagating systematic offsets between heterogeneous literature sources into the posterior.

\subsubsection*{Metallicity}
Metallicity enters the model through the initial metallicity $Z_\mathrm{ini}$, derived from [Fe/H] via
\begin{equation}
    Z_\mathrm{ini} = Z_\odot \times 10^{[\mathrm{Fe/H}]},
    \label{eq:feh2z}
\end{equation}
with the PARSEC solar metallicity $Z_\odot = 0.0152$. For clusters with at least three probable members that have Gaia XP spectrophotometric metallicity estimates ($N = 1\,201$ in our sample), we place a truncated Gaussian prior
\begin{equation}
    p([\mathrm{Fe/H}]) \propto \mathcal{N}([\mathrm{Fe/H}]_\mathrm{XP},\,0.3^2),
\end{equation}
truncated at $-1.5 \leq [\mathrm{Fe/H}] \leq +0.5$.  The choice of a 0.3 dex standard deviation serves as a weakly informative prior. While individual Gaia DR3 XP metallicity estimates typically report formal uncertainties of ~0.1-0.2 dex \citep[e.g.][]{Andrae2023}, larger systematic offsets can arise due to crowding, high extinction, or low signal-to-noise ratios in fainter cluster members. A broad scatter of 0.3 dex conservatively accounts for these systematics. More importantly, in our Bayesian framework, this width restricts the nested sampler to the physically relevant parameter space without overly constraining it, allowing the CMD likelihood to dominate the final posterior and pull the solution away from the XP prior mean if the photometric evidence demands it.

For the remaining clusters, we adopt a uniform prior whose range reflects the expected Galactic environment: $\mathcal{U}(-0.5, +0.5)$ for the standard disc, $\mathcal{U}(-1.5, 0.0)$ for the outer disc ($d > 15$\,kpc or $|b| > 30\degr$), and $\mathcal{U}(-0.3, +0.5)$ for the inner disc ($R_\mathrm{GC} < 5$\,kpc).

The nested sampler explores the full prior volume; if the CMD likelihood requires a metallicity far from the XP mean, the posterior will shift accordingly. Clusters for which the photometric CMD geometry provides little discriminating power will return a broad metallicity posterior — this is a faithful representation of the available information, not a failure of the method, and such cases are identifiable through the $\eta_\mathrm{fit}$ quality metric (Section~\ref{sec:quality}).

\subsubsection*{Distance}
The distance prior is a truncated Gaussian centred on the distance inferred from the median Gaia DR3 parallax of probable members,
\begin{equation}
    p(d) \propto \mathcal{N}(d_\varpi,\,\sigma_{d,\varpi}^2) \quad (d > 0),
\end{equation}
where $d_\varpi$ is the median parallax-inverted distance of the $N_\varpi$
probable members with reliable parallaxes. The prior constrains a single
quantity --- the characteristic distance of the cluster --- rather than the
distances of its individual stars. Its width is therefore set by the precision
with which that central distance is determined,
\begin{equation}
    \sigma_{d,\varpi} = \mathrm{clip}\!\left(
    \frac{1.4826\,\mathrm{MAD}\!\left[1000/\varpi_i\right]}{\sqrt{N_\varpi}},\;
    0.02\,d_\varpi,\; 0.50\,d_\varpi\right),
    \label{eq:distsigma}
\end{equation}
where the median absolute deviation (MAD) of the member parallax-inverted
distances $1000/\varpi_i$, scaled by $1.4826$ to approximate a Gaussian
dispersion, measures the observed scatter of member distances. Beyond a few
kiloparsecs this scatter is dominated by per-star parallax noise rather than by
the physical line-of-sight depth of the cluster (which is only of order tens of
parsecs); using it directly would therefore over-broaden the prior with
measurement noise. Because the prior pertains to the cluster's mean distance, we
scale the member scatter by $1/\sqrt{N_\varpi}$ so that the width reflects the
uncertainty on the central estimate rather than the dispersion of the individual
members. The result is bounded below at $2$~per~cent and above at
$50$~per~cent of $d_\varpi$; the upper bound prevents members with near-zero
parallax from inflating the width, and the hard boundary at $d = 0$ excludes
unphysical solutions. For the majority of clusters --- particularly rich, nearby
ones --- the $2$~per~cent lower clip sets the prior width, so the catalogue
distances are effectively \emph{astrometry-dominated}: the parallax anchor,
rather than the CMD likelihood, determines the reported distance and its
uncertainty. We caution that the $1/\sqrt{N_\varpi}$ reduction assumes
independent member parallaxes, whereas \textit{Gaia}~DR3 parallaxes carry a
spatially-correlated systematic ($\sim$$10$--$30\,\mu$as on the sub-degree
angular scale of a cluster; \citealt{Lindegren2021}) that does not average down.
The quoted $2$~per~cent uncertainties should therefore be regarded as a floor
that may be optimistic in the outer disc, and users requiring rigorous distance
error budgets should adopt full geometric distance priors
\citep[e.g.][]{BailerJones2021}.

We use the raw Gaia DR3 parallaxes without applying the
source-specific zero-point correction of \citet{Lindegren2021}, which
depends on each star's $G$ magnitude, colour, and ecliptic latitude and
therefore differs from star to star within a cluster.  Applying this
correction robustly to a cluster median parallax would require
propagating star-by-star corrections — a step we defer to future work.
The global-average Lindegren correction is $\approx -17\,\mu$as (the
corrected parallax is larger, shifting $d_\varpi$ to smaller values),
which corresponds to a distance shift of $\sim$7\,pc at 2\,kpc and
$\sim$170\,pc at 10\,kpc.  This uncorrected zero-point effect partially contributes to the small
negative distance offset ($\Delta d = X_\mathrm{ref} - X_\mathrm{this\,work} < 0$,
i.e.\ our distances are systematically larger than the literature) seen
in the literature comparisons (Section~\ref{sec:comparison:dist}).

\begin{figure*}
  \centering
  \includegraphics[width=\textwidth]{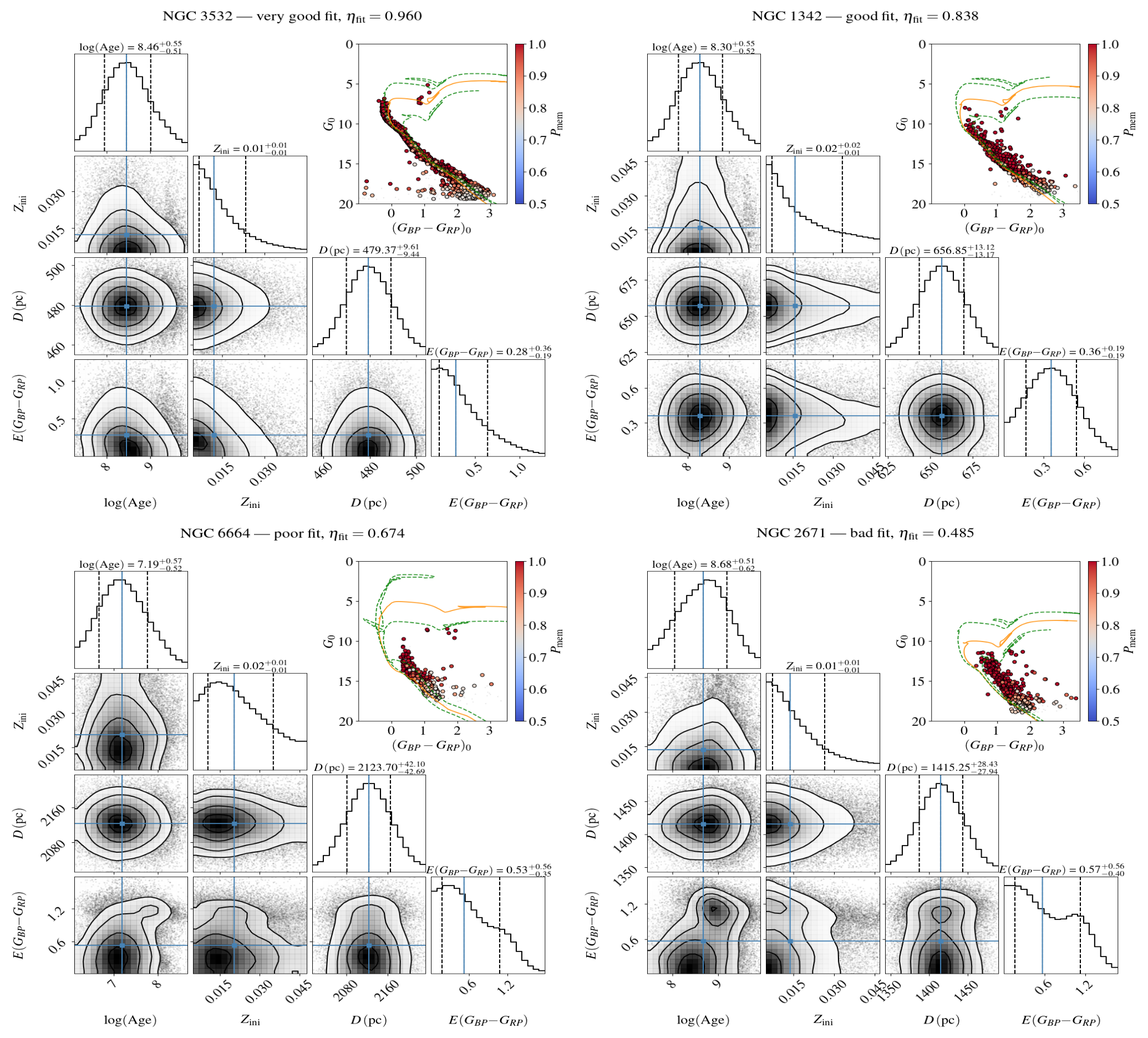}
  \caption{Posterior distributions (corner plots) for four clusters spanning
           the full quality range of $\etafit$.
           Each panel shows the joint and marginal posteriors for the four free
           parameters: $\log(\mathrm{Age/yr})$, initial metallicity $Z_\mathrm{ini}$,
           heliocentric distance $D$ (pc), and colour excess
           $E(G_\mathrm{BP}{-}G_\mathrm{RP})$; vertical dashed lines mark the
           16th, 50th, and 84th percentiles.
           The inset CMD in each panel shows stars colour-coded by membership
           probability (coolwarm scale), with the best-fitting PARSEC isochrone
           (orange) overlaid on the dereddened apparent-magnitude CMD.
           High-quality fits (NGC\,3532 at $\etafit = 0.960$ and NGC\,1342 at
           $0.838$) have compact, unimodal posteriors;
           lower-quality fits (NGC\,6664 at $0.674$ and NGC\,2671 at $0.485$)
           exhibit broader posteriors and elongated contours
           consistent with CMD degeneracies, not pipeline failure.}
  \label{fig:etafit_corners}
\end{figure*}

\subsubsection*{Reddening}
The reddening prior is a truncated Gaussian,
\begin{equation}
    p(E(G_{BP}-G_{RP})) \propto \mathcal{N}(\mu_E,\,\sigma_E^2) \quad (E(G_{BP}-G_{RP}) \geq 0),
\end{equation}
where the prior mean is
\begin{equation}
    \mu_E = K_{E(G_{\mathrm{BP}}{-}G_{\mathrm{RP}})/E(B-V)} \times E(B-V)_\mathrm{SFD} \times f_\mathrm{dist},
\end{equation}
using the \citet{Schlegel1998} full-sky dust map recalibrated by \citet{Schlafly2011}, and the conversion factor $K_{E(G_{\mathrm{BP}}{-}G_{\mathrm{RP}})/E(B-V)} = 1.305$ from \citet{Wang2019}. The factor $f_\mathrm{dist} \leq 1$ is a distance-dependent attenuation fraction that accounts for the fraction of the total line-of-sight dust column intercepted at the cluster distance.  It is computed from an exponential dust-disc model following \citet{Drimmel2001}:
\begin{equation}
  f_\mathrm{dist} = 1 - \exp\!\left(-\frac{d\,|\sin b|}{H_\mathrm{dust}}\right),
\end{equation}
where $d$ is the heliocentric distance (pc), $b$ is the Galactic latitude, and
$H_\mathrm{dust} = 125$\,pc is the exponential scale height of the thin dust
disc \citep{Drimmel2001}.  For clusters well above the plane
($d\,|\sin b| \gg H_\mathrm{dust}$) the factor approaches unity and the full
SFD column is used; for nearby low-latitude clusters it falls below unity,
preventing the integrated line-of-sight reddening from over-constraining the
foreground reddening.

The width of the reddening prior depends on Galactic latitude, reflecting the
larger uncertainty of the dust column towards the Galactic plane: we adopt
$\sigma_E = 0.50$\,mag for $|b| < 5\degr$, $\sigma_E = 0.35$\,mag for
$5\degr \leq |b| < 15\degr$, and $\sigma_E = 0.20$\,mag for $|b| \geq 15\degr$.
These three latitude bins contain $3\,988$, $849$, and $219$ of the $5\,056$
clusters, respectively.

The extinction in the $G$ band is computed as $A_G = K_{A_G/E(G_{\mathrm{BP}}{-}G_{\mathrm{RP}})} \times E(G_{BP}-G_{RP})$ with $K_{A_G/E(G_{\mathrm{BP}}{-}G_{\mathrm{RP}})} = 1.890$. The corresponding visual extinction is obtained via $A_V = A_G / 0.796$, adopting $R_V = 3.1$ and the \citet{Wang2019} extinction coefficients.

\begin{figure*}
  \centering
  \includegraphics[width=\textwidth]{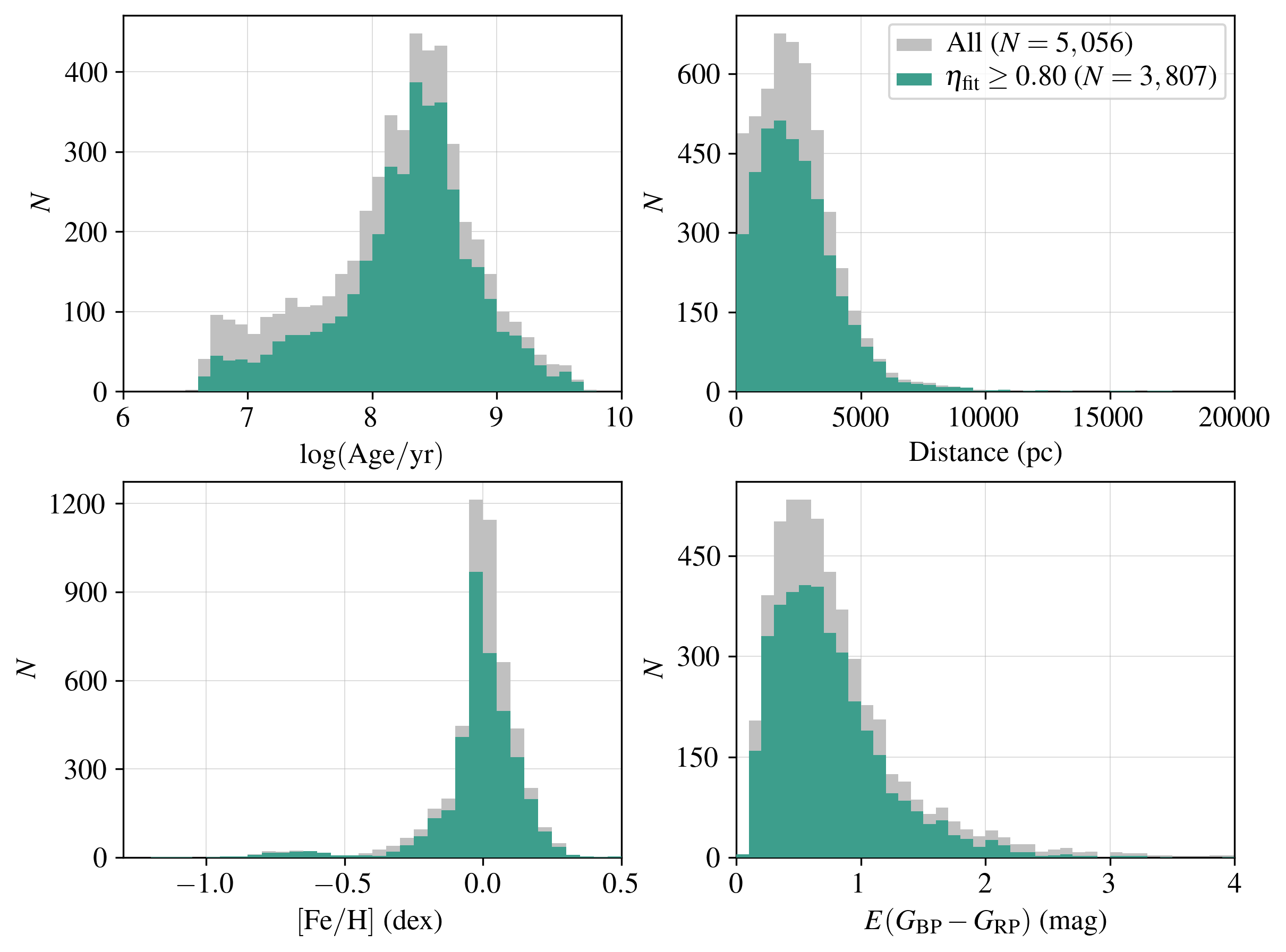}
  \caption{Distributions of the four derived parameters for the full sample
           (grey) and the high-quality subset ($\etafit \geq 0.80$, teal).
           \textit{Top left:} Logarithmic age.
           \textit{Top right:} Heliocentric distance.
           \textit{Bottom left:} Metallicity \feh.
           \textit{Bottom right:} Reddening in the Gaia bands, $E(G_{BP}-G_{RP})$.}
  \label{fig:params_dist}
\end{figure*}

\subsection{Likelihood function}
\label{sec:likelihood}

We model the CMD residuals as a weighted chi-squared statistic. For a given parameter vector $\boldsymbol{\theta}$, each probable member star $i$ (membership probability $p_i \geq 0.75$) is assigned the perpendicular distance $\Delta m_i$ to the nearest point on the dereddened, distance-shifted isochrone in the CMD plane, where the colour axis is $(G_\mathrm{BP} - G_\mathrm{RP}) - K_{E(G_{\mathrm{BP}}{-}G_{\mathrm{RP}})} \times E(G_{BP}-G_{RP})$ and the magnitude axis is $G - A_G - 5\log_{10}(d/10\,\mathrm{pc})$.  The nearest isochrone point is located by minimising the Euclidean distance in this two-dimensional magnitude space; since both axes carry units of magnitudes, the search assigns equal metric weight to colour and magnitude displacements, with the $\sigma_i$ normalisation entering solely through the likelihood denominator below. The log-likelihood is
\begin{equation}
    \ln \mathcal{L}(\boldsymbol{\theta}) = -\frac{1}{2} \sum_i w_i \left(\frac{\Delta m_i}{\sigma_i}\right)^2,
    \label{eq:lnL}
\end{equation}
where $w_i = p_i$ are the membership weights and $\sigma_i$ is the inflated photometric uncertainty,
\begin{equation}
    \sigma_i = \eta_\mathrm{infl} \sqrt{\sigma_{G,i}^2 + \min(\sigma_\mathrm{BP,i}^2,\,\sigma_\mathrm{RP,i}^2)},
\end{equation}
with an inflation factor $\eta_\mathrm{infl} = 1.5$.  This value was determined empirically by testing the pipeline on a set of well-studied clusters with independently known parameters, and was selected to yield well-behaved posteriors while absorbing the unmodelled scatter from differential reddening, unresolved binaries, and photometric systematics that are not explicitly captured by the isochrone model. We emphasise that the nearest-point construction makes this a \emph{pseudo}-$\chi^2$ statistic rather than a fully normalised generative likelihood: it omits the along-isochrone stellar-density (luminosity-function) weighting and does not model unresolved binaries explicitly. Accordingly, the reported log-evidence $\ln Z$ (tabulated for completeness) is meaningful only \emph{relatively}, as an internal fit-quality indicator, and should not be compared as a calibrated Bayesian evidence across clusters.

\subsection{Nested sampling implementation}
\label{sec:nested}

Posterior distributions are sampled with the \textsc{dynesty} package\footnote{\url{https://dynesty.readthedocs.io/}} \citep{Speagle2020}, which implements dynamic nested sampling \citep{Skilling2004} via its \textsc{DynamicNestedSampler} class. We initialise with $N_\mathrm{live} = 400$ live points and the MultiEllipsoid bounding method, which adapts to non-Gaussian and multi-modal posteriors. Sampling terminates when the estimated remaining log-evidence falls below $\Delta \ln Z = 0.01$. Each cluster is processed independently. Parameter estimates are reported as the posterior median together with the 16th and 84th percentile credible intervals ($\pm 1\sigma$ equivalent). The quality of the resulting posteriors across the full range of fit quality is examined in detail in Section~\ref{sec:results}.

\subsection{Quality metric \texorpdfstring{$\eta_\mathrm{fit}$}{eta\_fit}}
\label{sec:quality}

To assess the goodness of fit in a physically interpretable way, we define the \emph{inlier fraction} $\eta_\mathrm{fit}$: the fraction of probable members that fall within a $\pm 3\sigma$ band around the best-fitting isochrone (constructed using the posterior median parameters). Formally,
\begin{equation}
    \eta_\mathrm{fit} = \frac{1}{N_*} \sum_i 1\!\left[\Delta m_i \leq 3\sigma_i\right],
    \label{eq:etafit}
\end{equation}
where the sum runs over the $N_*$ stars passing the membership cut. A cluster is classified as a high-quality fit if $\eta_\mathrm{fit} \geq 0.80$; this threshold corresponds to a well-defined isochrone sequence cleanly separated from the field. Of the 5\,056 clusters in the final catalogue, 3\,807 (75.3\,per\,cent) satisfy this criterion.

Crucially, a broad posterior does not imply an absence of information: even when photometric constraints alone cannot pin down metallicity to better than $\sim$0.3\,dex, the derived posterior is still a proper, testable probability distribution over the physical parameter space. The $\eta_\mathrm{fit}$ metric captures \emph{convergence quality} — how well the best-fitting isochrone traces the observed CMD sequence — and shows no dependence on the widths of the adopted distance and reddening priors; clusters with higher $\eta_\mathrm{fit}$ nevertheless tend to have narrower posteriors in reddening and age, the parameters most directly constrained by the CMD. Clusters with low $\eta_\mathrm{fit}$ are retained in the published catalogue but flagged, allowing users to apply their own quality thresholds for population-level studies.

\subsection{Computational setup}
\label{sec:compute}

All calculations were performed on the TRUBA high-performance computing cluster operated by T\"{U}B\.{I}TAK ULAKBİM. Jobs were distributed via a SLURM array across 40 chunks for the AA-quality subsample (${\sim}55$ clusters per chunk) and 56 chunks for the AB-quality subsample (${\sim}59$ clusters per chunk). The average wall-clock time per cluster is 2--5\,min, depending on the number of member stars and the complexity of the CMD morphology. Clusters with a pre-existing output file are automatically skipped, making the pipeline fully restartable.
Representative examples of the CMD fits for four well-known clusters spanning
the full age range of the catalogue are shown in Fig.~\ref{fig:example_cmds}.

\begin{figure*}
  \centering
  \includegraphics[width=\textwidth]{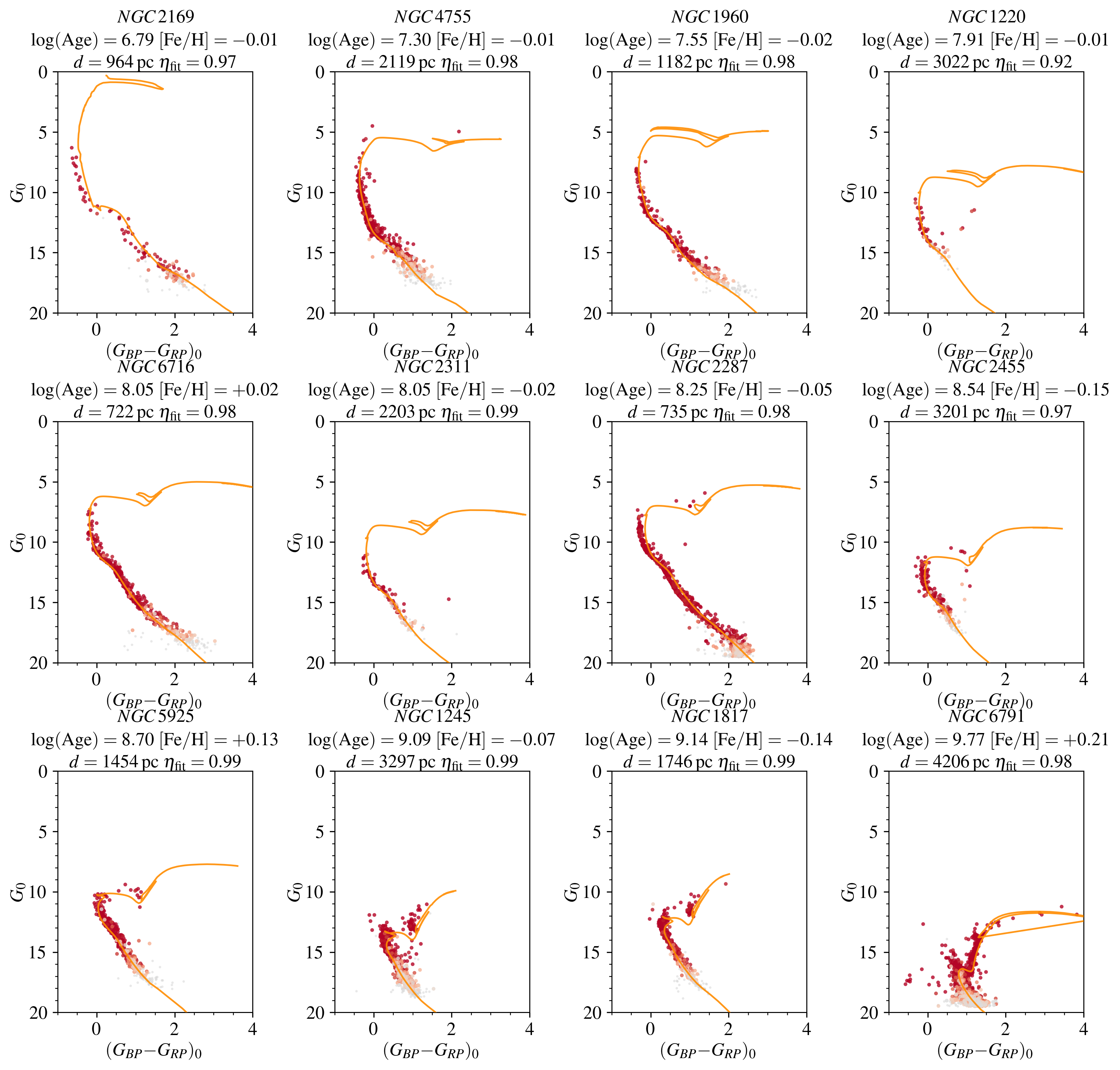}
  \caption{Dereddened colour--magnitude diagrams for 12 NGC open clusters
           with fit quality $\etafit \geq 0.90$, arranged in order of
           increasing age from top to bottom (three rows, four columns per row).
           Each panel shows probable members ($p \geq 0.75$) colour-coded by
           membership probability (coolwarm scale) with the best-fitting PARSEC
           isochrone overlaid (orange curve); grey points are stars below the
           membership threshold.
           The $y$-axis is the apparent dereddened magnitude $G_0 = G - A_G$
           and the $x$-axis is the dereddened colour $(G_\mathrm{BP}-G_\mathrm{RP})_0$.
           Annotated values give the derived $\log(\mathrm{Age})$, photometric \feh,
           heliocentric distance $d$, and inlier fraction $\etafit$.
           The sample spans from young clusters ($\sim$6\,Myr, top row)
           to old clusters (${\sim}6$\,Gyr, bottom row), demonstrating
           consistent isochrone fits across the full age baseline.}
  \label{fig:cmd_grid}
\end{figure*}
\begin{figure*}
  \centering
  \includegraphics[width=\textwidth]{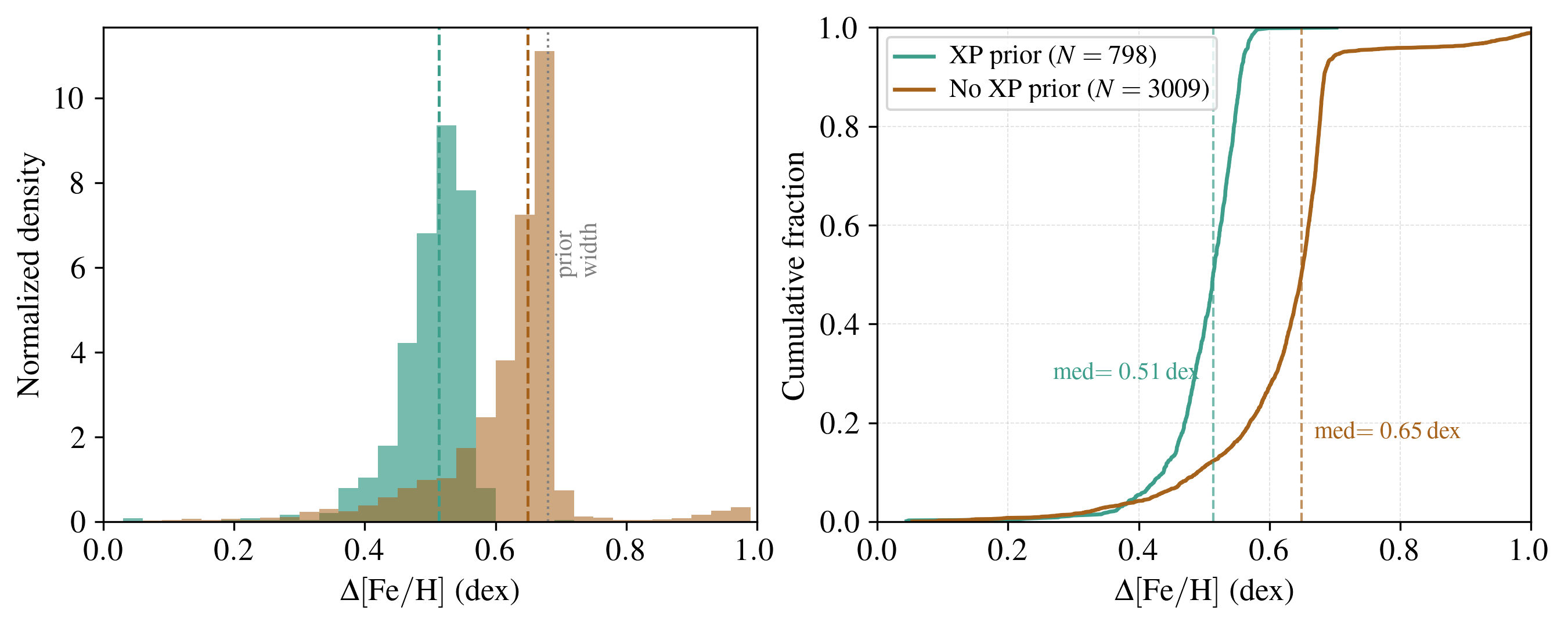}
  \caption{Comparison of \feh{} posterior widths
           ($\Delta[\mathrm{Fe/H}] \equiv [\mathrm{Fe/H}]_{84} - [\mathrm{Fe/H}]_{16}$)
           for clusters with (teal, $N = 798$) and without (brown, $N = 3\,009$)
           an informative \textit{Gaia} XP spectrophotometric metallicity prior,
           restricted to the high-quality subset ($\etafit \geq 0.80$).
           \textit{Left:} Normalized histogram; vertical dashed lines mark the
           respective medians.  The dotted vertical line at $0.68$\,dex indicates
           the $68$\% width of the broad uniform prior $[-0.5,+0.5]$\,dex
           adopted for clusters without XP coverage.
           \textit{Right:} Cumulative distribution functions.
           Clusters with XP priors have a median width of $0.51$\,dex,
           with $38.7$\% below $0.50$\,dex, whereas clusters
           without XP priors show a median of $0.65$\,dex ---
           comparable to the prior width itself --- reflecting the limited
           metallicity information available from the optical CMD alone.}
  \label{fig:feh_posterior_width}
\end{figure*}

\section{Results}
\label{sec:results}

\subsection{Overview of the catalogue}
\label{sec:results:overview}

We successfully processed all $5\,056$ AA and AB quality clusters in the UCC
\citep{Perren2023} that meet the minimum membership criterion
($2\,156$ class AA and $2\,900$ class AB; see Fig.~\ref{fig:spatial} for their
spatial distribution).
For each cluster, the Bayesian nested-sampling pipeline returns
posterior distributions for four parameters: $\log(\mathrm{Age/yr})$,
initial metallicity $Z_\mathrm{ini}$ (reported as [Fe/H]), heliocentric
distance $d$, and colour excess $E(G_\mathrm{BP}{-}G_\mathrm{RP})$.

Nested-sampling convergence (termination criterion $\Delta\ln Z < 0.01$)
was reached for $5\,036$ of the $5\,056$ clusters ($99.6$\,per\,cent);
the 20 non-converged clusters are flagged in the published catalogue.
The posterior log-evidence $\ln Z$ has a median of $-19.9$ and a range
of $-22.2$ to $-9.9$ across the full sample, consistent with well-behaved
likelihood surfaces for the great majority of clusters.

We adopt the \textit{fit quality indicator} $\eta_\mathrm{fit}$ to
separate reliable from unreliable fits.  Of the total sample,
$3\,807$ clusters ($75.3$\,per\,cent) satisfy $\eta_\mathrm{fit} \geq 0.80$
and form the high-quality subset used in all subsequent analyses.
The remaining $1\,249$ clusters either have poorly constrained posteriors
or show evidence of a degenerate CMD fit; these are retained in the
published catalogue but flagged accordingly.

The distribution of $\eta_{\mathrm{fit}}$ is shown in Fig.~\ref{fig:etafit_hist}.
The histogram peaks sharply near $\eta_{\mathrm{fit}} \approx 0.95$ and falls steeply
below $0.80$, indicating that the majority of the sample is well-fit and
that the adopted threshold separates a dense high-quality peak from a
low-quality tail rather than truncating a smooth distribution.
The four corner-plot panels in Fig.~\ref{fig:etafit_corners} confirm
that posterior quality tracks $\eta_{\mathrm{fit}}$ monotonically: high-quality
clusters (NGC\,3532 at $\eta_{\mathrm{fit}} = 0.960$; NGC\,1342 at $0.838$) show
compact, unimodal posteriors with tight credible intervals and isochrones
closely following the observed CMD sequences. Clusters near or below
the threshold (NGC\,6664 at $\eta_{\mathrm{fit}} = 0.674$; NGC\,2671 at $0.485$) exhibit
broader posteriors and elongated parameter contours, consistent with
the expected degeneracies in sparsely populated or highly reddened CMDs.
Critically, even these lower-quality fits do not show pathological
behaviour such as multi-modal posteriors or parameter values pegged
at prior boundaries; the quality reduction reflects a genuinely less
informative CMD rather than a pipeline failure.

For the high-quality subset the derived parameters span broad ranges
that reflect the diversity of the MW disc population.  Ages range from
$\log(\mathrm{Age/yr})$ = $6.60$ to $9.77$ (0.004--5.9\,Gyr), with a
median $\pm$ MAD of $8.35 \pm 0.30$ and a mean $\pm$ standard deviation
of $8.27 \pm 0.57$.  Heliocentric distances extend from 86 to
17\,012\,pc, with a median of 2\,172\,pc and a mean of 2\,444\,pc.
The [Fe/H] distribution spans $-1.17$ to $+0.46$\,dex, with a median
of $-0.001 \pm 0.061$\,dex (MAD).  The colour excess
$E(G_\mathrm{BP}{-}G_\mathrm{RP})$ ranges up to 3.914\,mag (equivalent
to $A_G \leq 7.40$\,mag), with medians of 0.658\,mag and 1.24\,mag
for $E(G_\mathrm{BP}{-}G_\mathrm{RP})$ and $A_G$, respectively.

Dividing the high-quality sample by age, we find 430 young clusters
($\log(\mathrm{Age/yr}) < 7.5$; 11.3 per cent), 2\,035 intermediate-age
clusters ($7.5 \leq \log(\mathrm{Age/yr}) < 8.5$; 53.5 per cent), and
1\,342 old clusters ($\log(\mathrm{Age/yr}) \geq 8.5$; 35.3 per cent).
The dominance of the intermediate-age group is consistent with
observational selection effects: very young clusters are often still
deeply embedded in their natal dust, while old clusters have had more
time to dissolve into the field.  The parameter distributions are
shown in Fig.~\ref{fig:params_dist}.
Fig.~\ref{fig:cmd_grid} presents CMD fits for a representative set of 12 NGC clusters selected to sample the full age baseline of the catalogue, covering six roughly equal logarithmic age bins from $\log(\mathrm{Age/yr}) \approx 6.79$ (${\sim}6$\,Myr) to $\log(\mathrm{Age/yr}) \approx 9.77$ (${\sim}5.9$\,Gyr); all 12 clusters satisfy $\eta_\mathrm{fit} \geq 0.90$. Each panel is plotted in dereddened apparent magnitude $G_0 = G - A_G$ versus intrinsic colour $(G_\mathrm{BP} - G_\mathrm{RP})_0$, with the orange curve showing the best-fitting PARSEC isochrone evaluated at the posterior median parameters.

The systematic evolution of CMD morphology with age is clearly visible. Young clusters display a luminous, vertically extended upper main sequence with no prominent red giant branch, while intermediate-age clusters exhibit a well-developed main-sequence turn-off point and a sparse red giant clump. Old clusters show a compact, faint turn-off together with a prominent red giant branch and clump — morphological hallmarks that are reproduced faithfully by the fitted isochrones. Across the full set of 12 clusters (Fig~\ref{fig:cmd_grid}) the inlier fraction $\eta_\mathrm{fit} \geq 0.90$, confirming that the isochrones trace the bulk of the observed sequences. The membership lists are nevertheless not free of contamination: a small number of photometric outliers, such as residual field stars and unresolved binaries, is visible in most CMDs (Figs~\ref{fig:example_cmds}, \ref{fig:etafit_corners} and \ref{fig:cmd_grid}). As these populations are not explicitly modelled in the likelihood, they can bias the inferred parameters, in particular towards higher extinction (Section~\ref{sec:comparison:dist}).

\subsection{Catalogue properties}
\label{sec:results:properties}

The median [Fe/H] of $-0.001$\,dex is indistinguishable from solar,
consistent with a sample dominated by disc clusters at Galactocentric
radii near the solar circle where the disc metallicity is close to the
solar value \citep{Casamiquela2021}.  The narrow MAD of 0.061\,dex reflects the informative starting point provided by Gaia\,XP photometric metallicities for clusters with sufficient member counts; for clusters without XP coverage, the environment-dependent uniform priors allow the CMD likelihood to drive the posterior across a much wider range. The full distribution consequently spans roughly 1.4\,dex, encompassing metal-poor clusters in the outer disc as well as moderately super-solar clusters near the Galactic bar.

To quantify the impact of the XP metallicity prior on posterior precision,
we compare the 68 per cent credible-interval widths
$\Delta[\mathrm{Fe/H}] \equiv [\mathrm{Fe/H}]_{84} - [\mathrm{Fe/H}]_{16}$
for the two groups within the high-quality subset
($\etafit \geq 0.80$; Fig.~\ref{fig:feh_posterior_width}).
Of the $1\,201$ clusters in our sample with an informative Gaia XP metallicity prior
(Section~\ref{sec:data:aux}), $798$ fall in the high-quality subsample and
form the XP group analysed here.  For these $798$ clusters the median
$\Delta[\mathrm{Fe/H}] = 0.51$\,dex, with $38.7$~per~cent of clusters
below $0.50$\,dex.  For the remaining $3\,009$ high-quality clusters that used
only the broad uniform prior $[-0.5,+0.5]$\,dex, the median width is
$0.65$\,dex --- comparable to the $0.68$\,dex 68~per~cent range of the
uniform prior itself --- and only $11.0$~per~cent have $\Delta[\mathrm{Fe/H}]
< 0.50$\,dex.  This demonstrates that, in the absence of spectrophotometric
constraints, the optical CMD alone provides limited metallicity discrimination
for the majority of clusters, and the reported [Fe/H] posteriors for the
non-XP subset should be interpreted as broad, CMD-modulated constraints
rather than precise individual measurements.  Users requiring tight
individual metallicity estimates for population-level chemical studies should
restrict their sample to the $798$ XP-prior clusters, or cross-match with
an independent spectroscopic survey such as OCCAM DR19
\citep{Otto2026}.  Across the full catalogue, the broad
posteriors are still statistically valid representations of the available
photometric information and remain useful for ensemble studies where
posterior widths can be propagated as measurement uncertainties.
For the $798$ high-quality XP-prior clusters, the posterior median [Fe/H]
differs from the XP prior centre by $+0.07$\,dex (median) with a scatter of
$0.10$\,dex.

The $E(G_\mathrm{BP}{-}G_\mathrm{RP})$ distribution is strongly
right-skewed. The sample median value for $E(G_\mathrm{BP}{-}G_\mathrm{RP})$ is $0.658$\,mag, but a significant tail extends to values above 2\,mag in Fig~\ref{fig:params_dist}.
These heavily reddened clusters are concentrated towards the Galactic
mid-plane (as visible in the Aitoff projection of Fig.~\ref{fig:spatial}),
where line-of-sight dust columns are highest.  Fitting isochrones to clusters with
$A_G \gtrsim 5$\,mag is inherently challenging because the CMD
becomes compressed and the blue main-sequence turn-off is no longer
accessible in optical passbands; the comparatively lower $\eta_\mathrm{fit}$
values for such objects reflect this difficulty.

\section{Comparison with literature catalogues}
\label{sec:comparison}

\begin{figure*}
  \centering
  \includegraphics[width=0.95\textwidth]{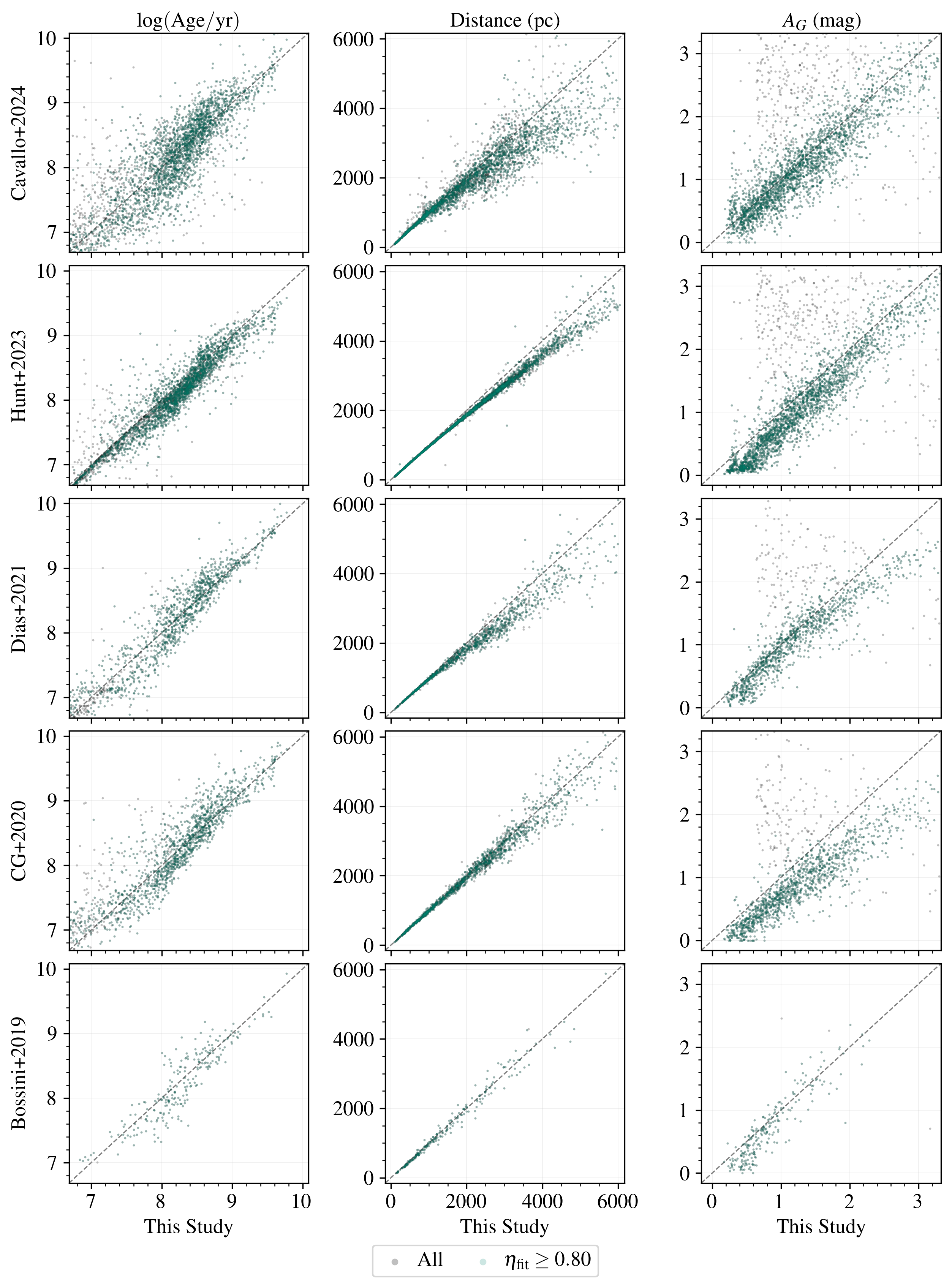}
  \caption{Scatter comparison of age ($\log(\mathrm{Age/yr})$), distance, and
           $A_G$ between this work and five photometric catalogues.
           Rows (top to bottom): Cavallo et~al.\ (2024), Hunt et~al.\ (2023),
           Dias et~al.\ (2021), Cantat-Gaudin et~al.\ (2020),
           Bossini et~al.\ (2019).
           Grey points show the full sample;
           teal points show clusters with $\etafit \geq 0.80$.
           Dashed lines indicate the 1:1 relation. The corresponding offset
           statistics are listed in Table~\ref{tab:comparison}.}
  \label{fig:compare_all4}
\end{figure*}

To validate our results we cross-match our catalogue against six independent
reference catalogues: \citet{Cavallo2024} (Cavallo+2024),
\citet{Hunt2023} (Hunt+2023),
\citet{Dias2021} (Dias+2021),
\citet{CantatGaudin2020} (Cantat-Gaudin+2020),
\citet{Bossini2019} (Bossini+2019), and the OCCAM
DR19 spectroscopic catalogue \citep{Otto2026}. Unless stated otherwise, the
statistics quoted in the text refer to the high-quality subset
($\eta_\mathrm{fit} \geq 0.80$). While
Sections~\ref{sec:comparison:cavallo24}--\ref{sec:comparison:bossini19} compare
against five photometric catalogues, the most stringent test of our
parameters is the independent spectroscopic metallicity cross-match with
OCCAM DR19 presented in Section~\ref{sec:comparison:occam}, whose agreement
underpins the metallicity-gradient application in
Section~\ref{sec:comparison:gradient}.
Throughout this section we define residuals as
$\Delta X = X_\mathrm{ref} - X_\mathrm{this\,work}$, and we quote
the mean $\pm$ standard deviation together with the median absolute
deviation (MAD) as a robust scatter estimate.
With this convention a \emph{negative} offset means that the value
derived in this work is larger than the reference value, while a
\emph{positive} offset means it is smaller.
A consolidated summary of all offset statistics, for both the high-quality
subset and all matched clusters, is given in
Table~\ref{tab:comparison}.  Scatter plots for the five photometric
catalogues are shown in Fig.~\ref{fig:compare_all4}, the corresponding
residuals as a function of our parameter values in Fig.~\ref{fig:compare_resid} (Appendix~\ref{app:resid}),
and the one-to-one OCCAM spectroscopic metallicity comparison in
Fig.~\ref{fig:compare_occam}. We note that four of the photometric reference
catalogues (Cavallo+2024, Hunt+2023, Dias+2021 and Cantat-Gaudin+2020) employ
synthetic CMDs as their fitting or training domain, in contrast to the analytic
PARSEC isochrone grid used here; this methodological difference may contribute
to the systematic offsets discussed below. \citet{Bossini2019}, in contrast,
fit PARSEC isochrones directly to \textit{Gaia}~DR2 $G$, $G_\mathrm{BP}$ and
$G_\mathrm{RP}$ photometry with the Bayesian MCMC code BASE-9, adopting
colour- and extinction-dependent extinction coefficients, a parallax-based
distance-modulus prior and a fixed, literature-based [Fe/H]; their approach is
therefore methodologically the closest to ours.

\subsection{Cavallo et al.\ (2024)}
\label{sec:comparison:cavallo24}

Cross-matching our catalogue with \citet{Cavallo2024} yields
$2\,922$ clusters in common; $2\,308$ satisfy our quality criterion
$\etafit \geq 0.80$.
\citet{Cavallo2024} estimated ages, metallicities, distances, and
extinctions for $\sim$5\,400 open clusters using an artificial
neural network (ANN) trained on synthetic clusters, with colour--magnitude
diagram features extracted via a QuadTree algorithm.
Their catalogue provides photometric \feh{} estimates for every cluster,
enabling a direct metallicity comparison --- an important cross-check
given that CMD-based metallicity recovery is one of the
principal goals of the present work.

For \logage{} we find
$\Delta\logage = -0.021 \pm 0.343$\,dex (MAD = $0.207$\,dex).
The negligible mean offset demonstrates that the two pipelines agree
on the age scale despite their fundamentally different fitting
strategies --- Bayesian nested sampling versus a QuadTree-feature ANN.
The scatter (MAD = $0.207$\,dex) is larger than that found for
all other compared catalogues,
which is expected given the deeper methodological differences:
our likelihood function explicitly traces isochrone sequences in CMD
space, whereas the ANN extracts topological features from the stellar
distribution without direct isochrone matching.

The metallicity comparison yields
$\Delta\feh = +0.060 \pm 0.396$\,dex (MAD = $0.218$\,dex).
The small positive offset indicates that our Bayesian estimates are
marginally lower on average than those of \citet{Cavallo2024}, while
the MAD is comparable to the
scatter found in the photometric comparison with Dias+2021
(Section~\ref{sec:comparison:dias21}).
The agreement is encouraging given that both studies rely solely on
Gaia broadband photometry and PARSEC isochrones yet derive metallicity
through entirely independent algorithms; the scatter of $\sim$0.4\,dex
is consistent with the intrinsic uncertainty of CMD-based
metallicity determinations reported in the literature
\citep[e.g.][]{Donor2020}.

Distances give $\Delta d = -274 \pm 579$\,pc (MAD = $153$\,pc),
consistent in sign with the offsets found for Hunt+2023 ($-179$\,pc),
Dias+2021 ($-224$\,pc), Cantat-Gaudin+2020 ($-44$\,pc), and Bossini+2019
($-21$\,pc).
Our pipeline systematically recovers larger distances than
\citet{Cavallo2024}, reinforcing the picture of a parallax-anchored
distance scale that lies above purely photometric estimates.
The \citet{Cavallo2024} ANN derives distance from the photometric
distance modulus alone, without an astrometric prior, making it
independently susceptible to photometric biases at large distances
where extinction uncertainty is high.
The larger scatter ($\sigma = 579$\,pc) relative to the other comparisons
reflects the increased methodological diversity between a
parallax-anchored Bayesian posterior and a purely photometric ANN
estimate.

For extinction we obtain $\Delta A_G = -0.108 \pm 0.329$\,mag
(MAD = $0.173$\,mag), where \citet{Cavallo2024} report $A_V$ values
that we convert to $A_G$ via $A_G = 0.796\,A_V$
(adopting $R_V = 3.1$ and the \citealt{Wang2019} coefficients).
The small negative mean offset indicates that our extinctions are
systematically somewhat higher than those of \citet{Cavallo2024}, in line
with the same-sign trend now seen against all five photometric catalogues
(Section~\ref{sec:comparison:dist}); the scatter remains comparable to the
other photometric comparisons.

\subsection{Hunt et al.\ (2023)}
\label{sec:comparison:hunt23}

Matching our catalogue to \citet{Hunt2023} yields
2\,337 clusters in common for $\etafit \geq 0.80$, making this the largest single overlap
in our validation set.  Hunt+2023 applied a neural-network isochrone
fitter to Gaia DR3 data using PARSEC models, providing ages,
distances, and extinctions; metallicity is not reported in that
catalogue, so the [Fe/H] comparison is omitted for this cross-match.

For $\log(\mathrm{Age/yr})$ we find
$\Delta\log(\mathrm{Age}) = -0.098 \pm 0.203$\,dex (MAD = 0.096\,dex).
The small negative offset indicates that our pipeline assigns slightly
older ages on average, while the MAD of 0.096\,dex demonstrates
excellent cluster-by-cluster agreement given that both studies use
PARSEC isochrones.  Residual systematic differences may arise from
differences in the CMD fitting strategy, the treatment of
differential reddening, and the way each method weights CMD
outliers.

Distances show $\Delta d = -179 \pm 285$\,pc (MAD = 101\,pc).  The
negative offset (i.e.\ our distances are larger) is plausibly related
to the deeper parallax-based distance prior adopted here, which anchors
the distance scale to the individual-star parallaxes of cluster members.  The scatter of
$\sim$300\,pc is broadly consistent with the photometric distance
uncertainties expected for clusters beyond $\sim$2\,kpc.

For extinction we obtain $\Delta A_G = -0.230 \pm 0.266$\,mag
(MAD = 0.133\,mag), where \citet{Hunt2023} report $A_V$ values that we
convert to $A_G$ via $A_G = 0.796\,A_V$ (adopting $R_V = 3.1$ and the
\citealt{Wang2019} coefficients).  The negative mean offset indicates that our
method recovers systematically higher extinctions than Hunt+2023 for matched
clusters, with a scatter ($\sigma = 0.266$\,mag) comparable to that of the other
comparisons, and in the same sense as all five photometric catalogues
(Section~\ref{sec:comparison:dist}).  This offset reflects the distinct
extinction \emph{definitions} rather than a zero-point error in either
pipeline: \citet{Hunt2023} report a mean $A_V$ together with a separate
differential term $\Delta A_V$, and explicitly document a
$\sim\!0.4$--$0.5$\,mag systematic between blue-edge and mean-extinction
definitions, whereas our pipeline fits the mean extinction over the full
member CMD (see the detailed discussion in Section~\ref{sec:comparison:dist}).

\begin{table*}
  \centering
  \caption{Summary of parameter offsets between this work and six reference
  catalogues, for the high-quality subset ($\etafit \geq 0.80$) and for all
  matched clusters. Residuals are
  defined as $\Delta X = X_\mathrm{ref}-X_\mathrm{this\,work}$;
  $\langle\Delta\rangle$ is the mean offset, $\sigma$ the standard deviation, and
  MAD the median absolute deviation from the median residual
  (Figs~\ref{fig:compare_all4} and \ref{fig:compare_resid}). All statistics are
  computed for clusters with valid values in both catalogues. $A_G$ values for this
  work are computed from the posterior median $\ebprp$ using
  $A_G = 1.890\,\ebprp$ \citep{Wang2019}; reference-catalogue extinctions
  reported as $A_V$ are converted via $A_G = 0.796\,A_V$, except for
  \citet{Bossini2019}, for which we adopt their tabulated turn-off $A_G$. OCCAM DR19
  \citep{Otto2026} provides MWM spectroscopic [Fe/H] and represents an
  independent spectroscopic validation of the photometric metallicity scale.}
  \label{tab:comparison}
  \begin{tabular}{llrrrrrrrr}
    \hline
    & & \multicolumn{4}{c}{$\etafit \geq 0.80$} & \multicolumn{4}{c}{All matched clusters} \\
    Reference & Parameter & $N$ & $\langle\Delta\rangle$ & $\sigma$ & MAD & $N$ & $\langle\Delta\rangle$ & $\sigma$ & MAD \\
    \hline
    OCCAM DR19 \citep{Otto2026} & $\Delta\feh$ (dex) & 132 & $-0.021$ & $0.132$ & $0.054$ & 141 & $-0.022$ & $0.129$ & $0.052$ \\
    \hline
    \citet{Cavallo2024} & $\Delta\logage$ (dex) & 2\,308 & $-0.021$ & $0.343$ & $0.207$ & 2\,922 & $-0.003$ & $0.410$ & $0.226$ \\
     & $\Delta d$ (pc) &  & $-274$ & $579$ & $153$ &  & $-269$ & $724$ & $185$ \\
     & $\Delta\feh$ (dex) &  & $+0.060$ & $0.396$ & $0.218$ &  & $+0.078$ & $0.436$ & $0.237$ \\
     & $\Delta A_G$ (mag) &  & $-0.108$ & $0.329$ & $0.173$ &  & $+0.005$ & $1.097$ & $0.234$ \\
    \citet{Hunt2023} & $\Delta\logage$ (dex) & 2\,337 & $-0.098$ & $0.203$ & $0.096$ & 2\,966 & $-0.088$ & $0.222$ & $0.090$ \\
     & $\Delta d$ (pc) &  & $-179$ & $285$ & $101$ &  & $-195$ & $374$ & $109$ \\
     & $\Delta A_G$ (mag) &  & $-0.230$ & $0.266$ & $0.133$ &  & $-0.084$ & $1.107$ & $0.185$ \\
    \citet{Dias2021} & $\Delta\logage$ (dex) & 1\,045 & $+0.007$ & $0.247$ & $0.167$ & 1\,217 & $+0.012$ & $0.254$ & $0.152$ \\
     & $\Delta d$ (pc) &  & $-224$ & $339$ & $123$ &  & $-228$ & $352$ & $128$ \\
     & $\Delta\feh$ (dex) &  & $+0.041$ & $0.148$ & $0.083$ &  & $+0.055$ & $0.160$ & $0.088$ \\
     & $\Delta A_G$ (mag) &  & $-0.160$ & $0.264$ & $0.138$ &  & $-0.149$ & $0.896$ & $0.164$ \\
    \citet{CantatGaudin2020} & $\Delta\logage$ (dex) & 1\,230 & $+0.022$ & $0.259$ & $0.154$ & 1\,437 & $+0.063$ & $0.307$ & $0.164$ \\
     & $\Delta d$ (pc) &  & $-44$ & $330$ & $57$ &  & $-44$ & $354$ & $65$ \\
     & $\Delta A_G$ (mag) &  & $-0.396$ & $0.280$ & $0.154$ &  & $-0.359$ & $0.873$ & $0.180$ \\
    \citet{Bossini2019} & $\Delta\logage$ (dex) & 229 & $-0.051$ & $0.242$ & $0.147$ & 232 & $-0.050$ & $0.241$ & $0.150$ \\
     & $\Delta d$ (pc) &  & $-21$ & $158$ & $46$ &  & $-19$ & $164$ & $46$ \\
     & $\Delta A_G$ (mag) &  & $-0.125$ & $0.213$ & $0.135$ &  & $-0.125$ & $0.285$ & $0.142$ \\
    \hline
  \end{tabular}
\end{table*}

\subsection{Dias et al.\ (2021)}
\label{sec:comparison:dias21}

The cross-match with \citet{Dias2021} provides 1\,045
common for $\etafit \geq 0.80$ clusters.  Dias+2021 derived parameters using PARSEC isochrones
with a Bayesian approach, and also provide [Fe/H] estimates based on
photometric membership.  Their metallicity priors incorporate a
position-dependent abundance--radius relation, in contrast to our
environment-conditioned uniform priors for clusters lacking XP
coverage.

For ages we find $\Delta\log(\mathrm{Age}) = +0.007 \pm 0.247$\,dex
(MAD = 0.167\,dex), indicating marginal agreement with a small
positive offset (i.e.\ a tendency for our ages to be younger).  The larger MAD relative to the
Hunt+2023 comparison (0.167 vs.\ 0.096\,dex) likely reflects the
heterogeneous nature of the Dias+2021 catalogue, which combines
literature values from many different sources, as well as differences
in the treatment of membership probabilities.

Distances give $\Delta d = -224 \pm 339$\,pc (MAD = 123\,pc).  The
negative offset is consistent with that found for Hunt+2023 and
reinforces the picture of a small but systematic distance scale
difference that warrants further investigation in future work.

The [Fe/H] comparison yields $\Delta[\mathrm{Fe/H}] = +0.041 \pm 0.148$\,dex
(MAD = 0.083\,dex).  The negligible mean offset and small MAD confirm
that our Bayesian metallicity estimates are in good overall agreement
with the photometric [Fe/H] values of Dias+2021.

For extinction we obtain $\Delta A_G = -0.160 \pm 0.264$\,mag
(MAD = 0.138\,mag), where \citet{Dias2021} report $A_V$ values that we
convert to $A_G$ via $A_G = 0.796\,A_V$ (adopting $R_V = 3.1$ and the
\citealt{Wang2019} coefficients).  The mean offset is negative,
indicating that our extinctions are systematically somewhat higher than those
of Dias+2021 consistent with the same-sign $A_G$ trend seen against
all five photometric catalogues (Section~\ref{sec:comparison:dist}); the
scatter is comparable to the other photometric comparisons.

\subsection{Cantat-Gaudin et al.\ (2020)}
\label{sec:comparison:cg20}

The overlap with \citet{CantatGaudin2020} comprises 1\,230 clusters for $\etafit \geq 0.80$.
Cantat-Gaudin+2020 used an artificial-neural-network (ANN) approach
trained on PARSEC isochrones to derive ages, distances, and extinctions
for clusters identified in Gaia DR2 data.

Ages show $\Delta\log(\mathrm{Age}) = +0.022 \pm 0.259$\,dex
(MAD = 0.154\,dex) consistent with the known limitations of ANN-based
fitting when applied to heterogeneous cluster CMDs and with the
improvement in member-star astrometry from DR2 to DR3.  The small
positive offset (i.e.\ our ages are slightly younger) mirrors that
found for Dias+2021 and suggests a modest
systematic trend that disappears when restricting the comparison to
well-populated clusters.

Distances yield $\Delta d = -44 \pm 330$\,pc (MAD = 57\,pc), the
smallest mean offset among the four catalogues based on synthetic CMDs; only
Bossini+2019 (Section~\ref{sec:comparison:bossini19}) shows a smaller offset.
The low MAD of 57\,pc indicates that, at the distance scale sampled by
Cantat-Gaudin+2020 (predominantly $d \lesssim 5$\,kpc), both methods
agree closely.

For extinction we find $\Delta A_G = -0.396 \pm 0.280$\,mag
(MAD = 0.154\,mag).  This is the largest extinction offset of the five
photometric comparisons, and its negative sign indicates that our
pipeline recovers systematically higher extinctions than
Cantat-Gaudin+2020 for matched clusters.  The magnitude of this offset
is driven primarily by their extinction \emph{definition}:
\citet{CantatGaudin2020} parametrise reddening as the monochromatic
extinction $A_0$ \citep{Danielski2018} and their fits effectively trace
the blue edge of the cluster sequence, placing their values
$\sim\!0.3$--$0.4$\,mag below the mean-extinction scale on which our
full-CMD fits operate (Section~\ref{sec:comparison:dist}).  The large
scatter is consistent with the difficulty of recovering extinction from
broad-band optical photometry alone.

All five photometric comparisons are displayed together in the
$5 \times 3$ scatter-plot grid of Fig.~\ref{fig:compare_all4}, with the
corresponding residuals in Fig.~\ref{fig:compare_resid} (Appendix~\ref{app:resid}). All offsets are listed in Table~\ref{tab:comparison}.

\subsection{Bossini et al.\ (2019)}
\label{sec:comparison:bossini19}

\citet{Bossini2019} derived ages, distance moduli, and extinctions for
269 open clusters from \textit{Gaia}~DR2 $G$, $G_\mathrm{BP}$, and
$G_\mathrm{RP}$ photometry by fitting PARSEC isochrones with the Bayesian code
BASE-9. Their sample is restricted to low-extinction ($A_V < 2.5$\,mag) clusters
older than 10\,Myr and located within $\sim$4.5\,kpc, and [Fe/H] is kept fixed at
literature values during the fit. The cross-match yields 232 clusters in
common, of which 229 satisfy $\etafit \geq 0.80$.

For ages we find $\Delta\logage = -0.051 \pm 0.242$\,dex
(MAD = 0.147\,dex), i.e.\ our ages are slightly older on average, in the same
sense as found for Hunt+2023, with a scatter similar to those of the Dias+2021
and Cantat-Gaudin+2020 comparisons.

Distances give $\Delta d = -21 \pm 158$\,pc (MAD = 46\,pc), the smallest
mean offset and scatter of all comparisons in this work. This close agreement
is partly expected: the matched sample is nearby (median distance
$\sim$1.1\,kpc; 43~per~cent within 1\,kpc), where parallax-based distances are
most reliable, and \citet{Bossini2019} fixed the distance modulus to the
parallax-inverted value for nearby ($d < 1$\,kpc), low-extinction clusters, so
that for part of the sample both analyses share a similar astrometric anchor.

For extinction we adopt the turn-off $G$-band extinction tabulated by
\citet{Bossini2019}, which they derive with colour- and extinction-dependent
\textit{Gaia} extinction coefficients \citep{Babusiaux2018}, and obtain
$\Delta A_G = -0.125 \pm 0.213$\,mag (MAD = 0.135\,mag). The negative offset
is in the same sense as for the other four photometric catalogues, with the
smallest scatter among them. As their extinction is evaluated at the turn-off,
whereas our fits trace the mean extinction over the full member CMD, a modest
offset in this direction is expected (Section~\ref{sec:comparison:dist}).

\subsection{Discussion of systematic offsets in distance and extinction}
\label{sec:comparison:dist}

Across all five photometric catalogues a consistent negative
distance offset is present: $-21$, $-44$, $-224$, $-179$ and $-274$\,pc
for Bossini+2019, Cantat-Gaudin+2020, Dias+2021, Hunt+2023 and Cavallo+2024,
respectively. The direction is the same in every case, suggesting a systematic rather
than random origin.
The most likely contributor is the depth of the parallax-based distance
prior adopted here: our Gaussian prior centred on the median parallax-inverted
distance ($\mu = \mathrm{median}(1000/\varpi_i)$) provides a strong anchor that
can shift the posterior towards larger values when the photometric
likelihood is broad.
The offset for Cavallo+2024 ($-274$\,pc) is the largest of the
five, consistent with the fact that their ANN derives distance purely
from the photometric distance modulus without any astrometric
constraint, leaving the posterior more susceptible to photometric
biases.
Parallax-inverted distances are known to be biased to larger values
for clusters at $d \gtrsim 1$\,kpc where the relative parallax error
exceeds $\sim 10$~per~cent
\citep{Lindegren2018}: the inversion of a noisy positive parallax
over-estimates the true distance (Lutz--Kelker bias).  We therefore expect
a mild negative bias to propagate into our distance posteriors and into
the offsets observed here.  The bias diminishes at close distances,
consistent with the notably smaller offsets seen against Cantat-Gaudin+2020
($-44$\,pc, MAD $= 57$\,pc), whose sample is dominated by clusters
within $5$\,kpc, and against Bossini+2019 ($-21$\,pc, MAD $= 46$\,pc), whose
matched sample has a median distance of $\sim$1.1\,kpc
(Section~\ref{sec:comparison:bossini19}).  Future work incorporating full geometric distance priors
\citep[e.g.][]{BailerJones2021} in place of the parallax-inverted
Gaussian will mitigate this effect. The residuals in Fig.~\ref{fig:compare_resid}
(Appendix~\ref{app:resid}) show that the distance offsets relative to Hunt+2023,
Dias+2021 and Cavallo+2024 grow with distance, consistent with this
interpretation, whereas no such trend is seen against Bossini+2019, whose
matched sample is predominantly nearby.

The mean $A_G$ offsets across the five photometric catalogues
($-0.125$, $-0.396$, $-0.160$, $-0.230$ and $-0.108$\,mag for
Bossini+2019, Cantat-Gaudin+2020, Dias+2021, Hunt+2023 and Cavallo+2024, respectively) all
share the same sign: our extinctions are modestly but systematically higher than
those of the reference catalogues. Their magnitude, however, is strongly
catalogue-dependent, and this dependence points to the definition and
parametrisation of extinction --- rather than a zero-point error in our
posteriors --- as the dominant cause.

We first rule out the conversion coefficient as the source of the trend. Our
extinctions adopt $A_G / A_V = 0.796$ \citep{Wang2019}; evaluating instead the
full colour- and extinction-dependent \textit{Gaia} $G$-band extinction
polynomial \citep{Babusiaux2018} at a representative cluster main-sequence
colour yields $k_G = A_G/A_0 \simeq 0.78$, i.e.\ agreement with our constant
coefficient to within $2$~per~cent. Replacing the constant factor by the
per-cluster polynomial changes the offsets by $<0.06$\,mag and does \emph{not}
reduce them, confirming that the systematic is not a conversion artefact.

Instead, the offsets track the heterogeneous ways in which the reference
catalogues define and measure extinction. \citet{CantatGaudin2020}
parametrise reddening as the monochromatic extinction $A_0$ using the
\textit{Gaia} extinction polynomial of \citet{Danielski2018}, and their
neural-network fits effectively trace the blue edge of the cluster sequence;
on the matched high-quality sample their values lie systematically
$\sim\!0.3$--$0.4$\,mag below those of \citet{Hunt2023} and \citet{Dias2021} on
a common $A_V$ scale (matched medians $1.00$ versus $1.29$ and $1.35$\,mag),
which naturally makes their $A_G$ offset the largest of the five.
\citet{Hunt2023} report a mean extinction $A_V$ together with a
\emph{separate} differential term $\Delta A_V$, and explicitly document a
$\sim\!0.4$--$0.5$\,mag systematic between blue-edge and mean-extinction
definitions, as well as a $\sim\!0.4$\,mag overestimate of $\Delta A_V$ induced
by unresolved binaries. \citet{Dias2021} fit $A_V$ using the same colour- and
extinction-dependent \textit{Gaia} polynomial \citep{Monteiro2020} together
with a three-dimensional dust-map prior, and show the closest agreement with
our values. \citet{Cavallo2024}, the most homogeneous \textit{Gaia}~DR3 fit of
the four synthetic-CMD-based catalogues, exhibits the smallest offset.
\citet{Bossini2019} quote the extinction at the cluster turn-off, derived with
colour- and extinction-dependent coefficients. Because our pipeline fits the full
member CMD --- and therefore the mean extinction averaged over all probable
members --- it is expected to lie above blue-edge- and turn-off-based estimates; an
asymmetric colour-residual likelihood or an explicit binary-sequence component
could anchor the fitted isochrone to the blue edge in future work. All five
offsets remain within $\lesssim 2.6$ times the corresponding MAD scatter, and
are therefore consistent with definitional differences rather than a gross
extinction bias in our posteriors.

Superimposed on this definitional trend, inspection of the scatter plots
(Fig.~\ref{fig:compare_all4}) reveals a distinct sub-population of heavily
reddened clusters for which our pipeline instead recovers near-zero $A_G$ while
the literature assigns large extinctions. For clusters embedded in or behind
dense dust clouds, the finite-distance-corrected SFD column can underestimate
the true foreground, allowing the reddening prior to pull the posterior towards
lower values even when the CMD morphology would support higher reddening; this
is compounded by the distance--extinction covariance inherent in isochrone
fitting, whereby a larger (parallax-anchored) distance posterior is partially
absorbed by a lower extinction. A comparable anti-correlation between $A_V$ and
distance was reported by \citet{Dias2021} relative to \citet{CantatGaudin2020}.

As an independent check of our extinction scale, we compared our reddenings
with the Combined19 three-dimensional dust map
\citep{Drimmel2003,Marshall2006,Green2019}, evaluated at the distance of each
cluster with \texttt{mwdust} \citep{Bovy2016}. For the high-quality subset the
agreement improves from the prior (MAD $= 0.28$\,mag in $\ebprp$) to the
posterior (MAD $= 0.16$\,mag), the latter being smaller than the differences
between the three-dimensional maps themselves (MAD $= 0.21$\,mag between
Combined19 and the map of \citealt{Drimmel2003}). Our posterior reddenings are,
however, systematically higher by $\sim$0.12\,mag in $\ebprp$, in the same sense
as the $A_G$ offsets found against the literature catalogues. Part of this excess
is likely due to photometric outliers --- residual field contamination and
unresolved binaries among the fainter members --- which are not explicitly
modelled in the likelihood and preferentially scatter stars redwards of the
single-star sequence: the excess tends to increase with the median magnitude of the members,
and clusters with a larger fraction of faint members show both a larger excess
and a lower $\etafit$. This effect accounts for only part of the offset.
Because of the extinction--distance--age correlations inherent in isochrone
fitting, this same outlier population contributes, to a lesser extent, to the
posteriors of the other parameters; this contribution is already reflected in
the reported credible intervals and in the lower $\etafit$ of the affected
clusters, rather than constituting an additional, unquantified source of
error.

\begin{figure}
  \centering
  \includegraphics[width=\columnwidth]{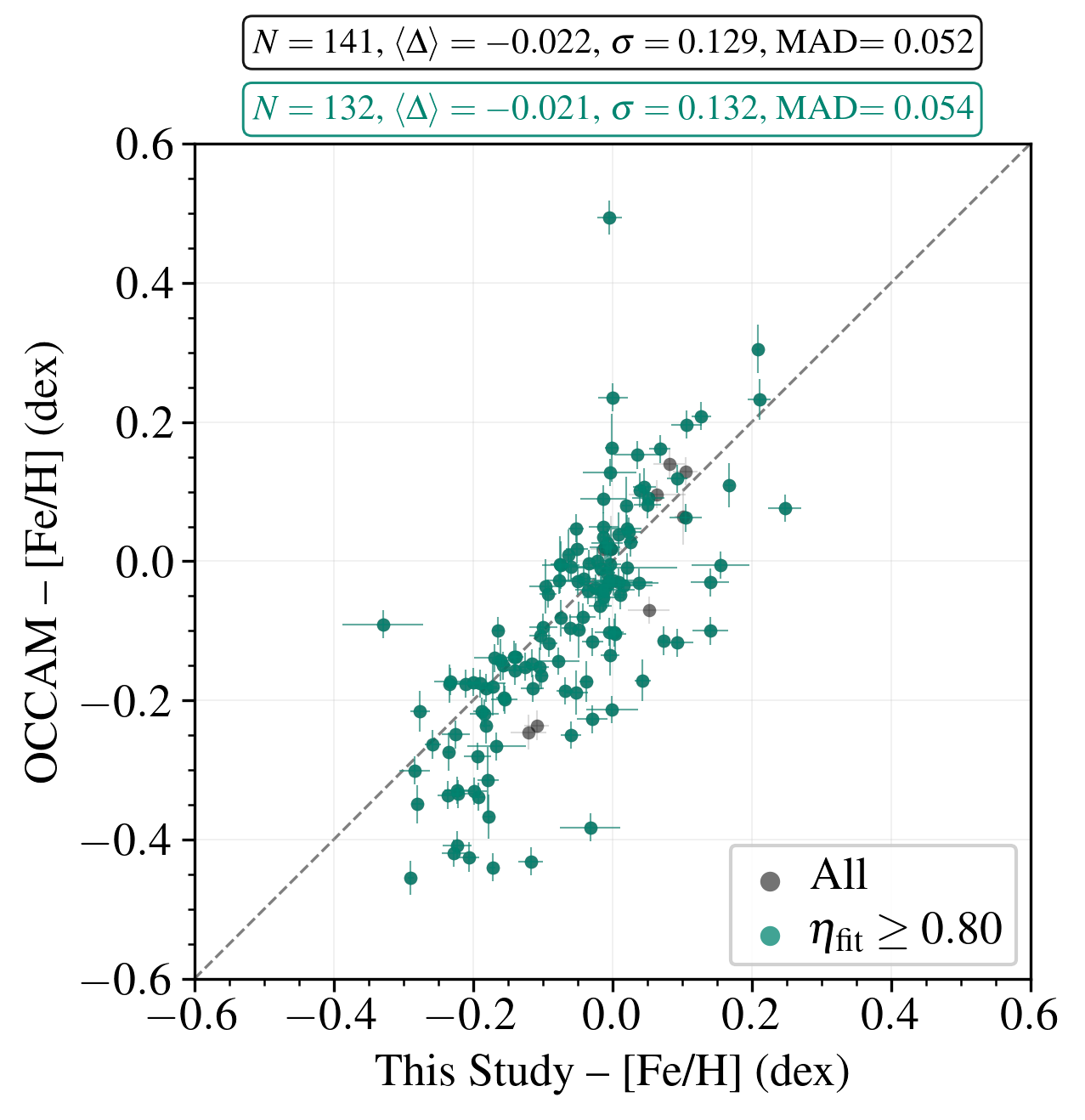}
  \caption{One-to-one comparison between photometric \feh{} (this work) and
           MWM spectroscopic \feh{} from the OCCAM DR19 catalogue
           \citep{Otto2026} for the $\etafit \geq 0.80$ subset
           (teal) and the full sample (grey).
           The dashed line shows the 1:1 relation.
           Statistics ($N$, mean offset, $\sigma$, MAD) are shown above the panel.
           Horizontal error bars show the standard error of our [Fe/H],
           $([\mathrm{Fe/H}]_{84}-[\mathrm{Fe/H}]_{16})/(2\sqrt{N_*})$, where
           $N_*$ is the number of members used in the fit; vertical error bars
           are the uncertainties reported in OCCAM DR19.}
  \label{fig:compare_occam}
\end{figure}

\subsection{OCCAM DR19 — spectroscopic [Fe/H]}
\label{sec:comparison:occam}

The most stringent test of our metallicity scale is provided by the
OCCAM DR19 catalogue \citep{Otto2026}, which lists
high-resolution Milky Way Mapper \citep[MWM:][]{Kollmeier2026} spectroscopic [Fe/H] measurements for open
clusters in the MW disc.  We find 132 clusters in common after applying
the $\eta_\mathrm{fit} \geq 0.80$ quality cut.

The residuals are $\Delta[\mathrm{Fe/H}] = -0.021 \pm 0.132$\,dex
(MAD = 0.054\,dex), demonstrating excellent agreement between our
photometric-Bayesian metallicities and independent high-resolution
spectroscopy.  The mean offset of only $-0.021$\,dex is well within
the systematic uncertainty of isochrone-based metallicity determinations
($\lesssim 0.1$\,dex) and amounts to only ${\sim}0.2$ times the
cluster-to-cluster scatter.  The tight MAD of 0.054\,dex underscores the reliability of
the Gaia\,XP photometric prior combined with the nested-sampling posterior
in recovering [Fe/H] values that are close to spectroscopic ground
truth.

The 132 OCCAM clusters in common span Galactocentric radii
$R_\mathrm{GC} = 5.7$--$14.0$\,kpc (median $9.5$\,kpc; $R_0 = 8.2$\,kpc),
heliocentric distances of $86$--$7\,430$\,pc (median $2\,558$\,pc), and ages
$\log(\mathrm{Age/yr}) = 6.8$--$9.8$ (median $8.80$); $70$~per~cent
have $\log(\mathrm{Age/yr}) \geq 8.5$, confirming that the subsample
is dominated by intermediate-to-old disc clusters spanning a wide
range of Galactocentric radii \citep{Donor2020}.

To assess the role of the \textit{Gaia} XP metallicity prior in
the spectroscopic agreement, we split the OCCAM validation sample
into clusters with ($N = 89$) and without ($N = 43$) a Gaia\,XP-based Gaussian metallicity prior (see Section~\ref{sec:data:aux}).  For the XP-prior subsample the residuals are
$\Delta[\mathrm{Fe/H}] = -0.037 \pm 0.090$\,dex (MAD $= 0.056$\,dex),
while for the non-XP subsample they are
$+0.011 \pm 0.190$\,dex (MAD $= 0.065$\,dex).  Two findings emerge
from this decomposition.  First, both mean offsets remain small
($-0.037$ vs.\ $+0.011$\,dex, both $\lesssim 0.04$\,dex), confirming that neither group is
systematically biased with respect to MWM.  Second, the scatter
is substantially smaller for XP-prior clusters (std $0.090$ vs.\
$0.190$\,dex), consistent with the tighter posterior widths seen in
Fig.~\ref{fig:feh_posterior_width}.  Critically, even the non-XP
group has a scatter (MAD $= 0.065$\,dex) that is well below the
$0.66$\,dex typical posterior width for that group, confirming that
the CMD likelihood does constrain the metallicity despite the broad
prior — the posterior is not simply a copy of the prior, but the
constraint is weaker and the reported uncertainty faithfully reflects
this.  Users of the catalogue who require tight individual metallicity
estimates should preferentially select clusters with an informative
XP prior (flaggable via the cross-match with the Gaia\,XP catalogue).

The small MAD $= 0.054$\,dex across the full OCCAM sample — well within
the systematic uncertainty of isochrone-based metallicity determinations
($\lesssim 0.1$\,dex) — confirms that the photometric metallicity scale
of our catalogue is consistent with independent high-resolution
spectroscopy across the full range of ages, distances, and Galactic
positions represented in the validation sample.  The comparison is illustrated in Fig.~\ref{fig:compare_occam} (one-to-one plot with summary statistics).

\subsection{The radial metallicity gradient: a population-level test}
\label{sec:comparison:gradient}

\begin{figure*}
  \centering
  \includegraphics[width=\textwidth]{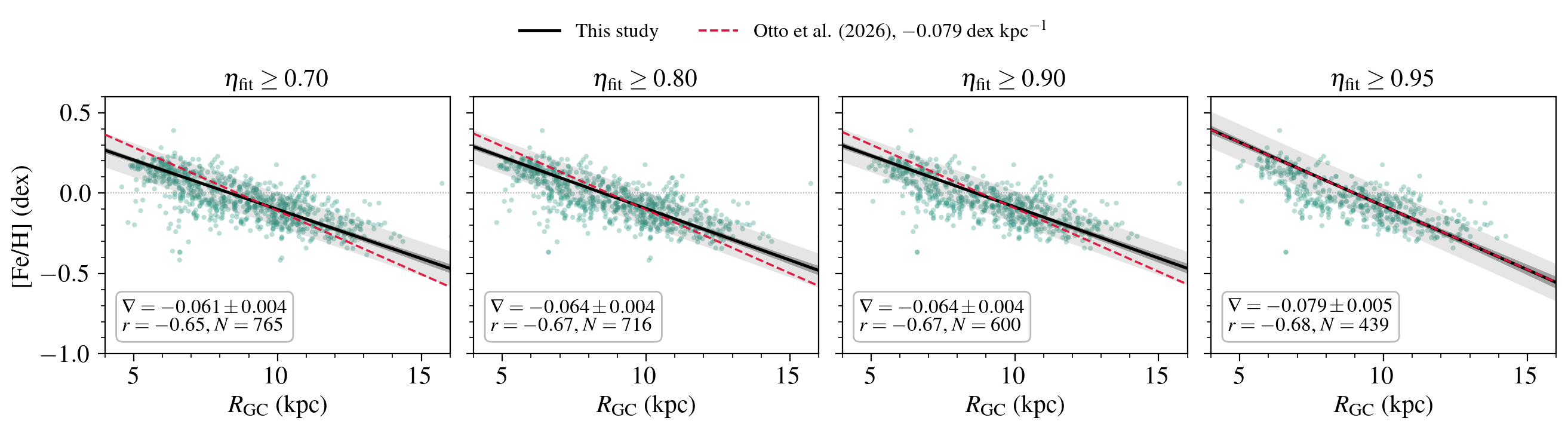}
  \caption{Radial metallicity gradient for the XP-prior clusters at four
  fit-quality thresholds ($\eta_\mathrm{fit}\geq0.70$--$0.95$, left to right).
  Teal points are individual clusters (all XP-prior clusters passing each cut are
  shown; the linear fit is performed over $6$--$14$\,kpc and extrapolated across
  the plotted range). The black line is our weighted fit. \emph{Two shaded bands}
  are shown and should not be confused: the dark, narrow band is the $68$ per
  cent \emph{confidence interval on the mean relation} --- the uncertainty of the
  fitted line itself, which is small because the slope is determined by several
  hundred clusters --- while the light, wide band is the $68$ per cent
  \emph{prediction interval}, which additionally includes the intrinsic
  cluster-to-cluster scatter of the photometric metallicities and therefore
  traces the spread of the individual points. The red dashed line is the
  spectroscopic gradient of \citet{Otto2026} ($-0.079$\,dex\,kpc$^{-1}$); our fit
  steepens towards it as the quality cut is tightened, converging onto it at
  $\eta_\mathrm{fit}\geq0.95$.}
  \label{fig:gradient}
\end{figure*}

The radial metallicity gradient of the Galactic disc is one of the most
enduring observational constraints on models of disc formation and chemical
evolution. The thin disc exhibits a negative radial metallicity profile, with
the inner regions more metal-rich than the outer disc \citep{LuckLambert2011,
Huang2015, GaiaCollab2023b}, a signature also recovered by radial-gradient
measurements from field main-sequence and red-clump stars \citep{Plevne2015,
Akbaba2025}, and commonly interpreted as evidence of
inside-out disc growth \citep{Larson1976, Chiappini1997, Frankel2019} and
subsequently reshaped by radial migration \citep{SellwoodBinney2002, Minchev2013,
Spitoni2019, Akbaba2025}. Open clusters are premier tracers of this
gradient because their distances, ages, and abundances can all be determined
comparatively reliably and over a wide range of Galactocentric radius
\citep{Friel1995, Friel2002, Magrini2017, CantatGaudin2020}. Spectroscopic
surveys of open clusters have converged on an inner-disc slope of roughly
$-0.06$ to $-0.08$\,dex\,kpc$^{-1}$
\citep{Frinchaboy2013, Netopil2016, Casamiquela2019, Donor2020,
Spina2021, Spina2022, Myers2022, Magrini2023, Otto2026}, with photometric-metallicity
determinations yielding consistent values \citep{Netopil2016, Netopil2022}.
The behaviour beyond $R_\mathrm{GC}\simeq12$--$13$\,kpc remains debated, with
several studies reporting a flattening of the gradient in the outer disc
\citep{Yong2012, Frinchaboy2013, Donor2020} --- a regime in which
cluster distances and metallicities are also the hardest to measure. Our
homogeneous, free-metallicity catalogue offers an opportunity both to
demonstrate the catalogue's fitness for such studies and to provide a
population-level consistency check of our photometric metallicity scale that
complements the cluster-by-cluster OCCAM cross-match.

We compute the Galactocentric radius as
$R_\mathrm{GC} = (R_0^2 + (d\cos b)^2 - 2 R_0\, d\cos b \cos l)^{1/2}$ with
$R_0 = 8.2$\,kpc \citep{GravityCollab2019}, and fit a weighted linear relation
$\mathrm{[Fe/H]} = \nabla_{\!\feh}\,R_\mathrm{GC} + c$ over the well-populated
range $6 \le R_\mathrm{GC} \le 14$\,kpc, weighting each cluster by
$1/\sigma_\feh^2$. Because our metallicities are informative only for the XP
subset (Section~\ref{sec:results:properties}), we measure the gradient
separately for the full high-quality sample and for the XP-prior clusters, and
across a sequence of quality cuts $\eta_\mathrm{fit} \geq 0.70, 0.80, 0.90,
0.95$ (Table~\ref{tab:gradient}; Fig.~\ref{fig:gradient}).

The two subsets behave very differently. For the full high-quality sample the
recovered gradient is shallow and only weakly correlated
($\nabla_{\!\feh}\approx-0.05$\,dex\,kpc$^{-1}$, $r\approx-0.5$) and barely
changes as the quality cut is tightened. This is a direct consequence of prior
dominance: the $\sim$79 per cent of high-quality clusters without an XP prior
regress towards the near-solar prior mean (Section~\ref{sec:results:properties}),
diluting and flattening any underlying gradient.

Restricting the fit to the XP-prior clusters --- where $\feh$ carries genuine
\textit{Gaia} XP spectrophotometric information --- recovers a distinctly
steeper and tighter gradient that sharpens as the quality threshold increases:
$\nabla_{\!\feh} = -0.064 \pm 0.004$\,dex\,kpc$^{-1}$ ($r=-0.67$) at
$\eta_\mathrm{fit}\geq0.80$, steepening to $-0.079 \pm 0.005$\,dex\,kpc$^{-1}$
($r=-0.68$) at $\eta_\mathrm{fit}\geq0.95$. The latter is in remarkable
agreement with the spectroscopic gradient derived from the same OCCAM DR19
sample used above, $-0.079 \pm 0.006$\,dex\,kpc$^{-1}$ \citep{Otto2026}, and
sits squarely within the range of spectroscopic and photometric open-cluster
determinations \citep{Frinchaboy2013, Netopil2016, Casamiquela2019, Donor2020,
Spina2021, Myers2022, Magrini2023}. In particular, our $\eta_\mathrm{fit}\geq0.80$
XP slope of $-0.064 \pm 0.004$\,dex\,kpc$^{-1}$ coincides with the
$-0.064 \pm 0.007$\,dex\,kpc$^{-1}$ obtained by \citet{Spina2022} from
175 open clusters with high-quality spectroscopic metallicities. That our
cluster-based slope is somewhat steeper than radial gradients measured from
older field main-sequence and red-clump populations \citep{Plevne2015, Akbaba2025}
is consistent with the expectation that radial migration progressively flattens
the gradient traced by dynamically older stars \citep{Frankel2019, Akbaba2025},
whereas open clusters --- being on average younger and less radially mixed ---
retain a steeper, more nearly primordial imprint of the disc's chemical
enrichment.

This gradient analysis serves three purposes. It provides a population-level
consistency check of our photometric metallicity scale, complementary to the
cluster-by-cluster OCCAM comparison of Section~\ref{sec:comparison:occam}; we
stress, however, that because the XP-informed metallicities inherit the
\citet{Andrae2023} XP scale --- itself trained on APOGEE-family spectroscopic
labels, the same lineage as OCCAM~DR19 --- the recovered gradient demonstrates
that the pipeline faithfully \emph{preserves} the spectroscopically-encoded
gradient rather than deriving it independently of spectroscopy. It also
demonstrates the catalogue's readiness
for Galactic chemical-evolution applications; and it confirms operationally the
central framing of this work
(Sections~\ref{sec:intro},~\ref{sec:results:properties}) --- the true disc
gradient emerges only when the analysis is confined to the XP-informed clusters,
and it cannot be recovered by tightening the fit-quality cut alone, because the
dilution is set by metallicity \emph{information content} (the presence of an XP
prior) rather than by fit quality.

\section{Summary and Conclusions}
\label{sec:conclusions}

We have presented a homogeneous determination of fundamental astrophysical
parameters --- age, initial stellar metallicity ($Z_\mathrm{ini}$, converted to
\mbox{[Fe/H]}), heliocentric distance, and selective extinction $E(G_\mathrm{BP}{-}G_\mathrm{RP})$
--- for $5\,056$ open clusters drawn from the Unified Cluster Catalogue
\citep[UCC;][]{Perren2023}.  The analysis employs Bayesian Nested
Sampling \citep{Speagle2020} coupled with PARSEC isochrones
\citep{Bressan2012, Marigo2017} in the Gaia DR3 photometric
system \citep{GaiaCollab2023}, constituting one of the largest uniformly derived
open-cluster parameter catalogues to date.  The principal findings are
summarised below.

\begin{table}
  \centering
  \caption{Radial metallicity gradient $\nabla_{\!\feh}$ (dex\,kpc$^{-1}$) from a
  weighted linear fit over $6 \le R_\mathrm{GC} \le 14$\,kpc ($R_0 = 8.2$\,kpc),
  for the full high-quality sample and for the XP-prior clusters, at four
  fit-quality thresholds. $r$ is the Pearson correlation coefficient and $N$ the
  number of clusters in the fit. For reference, the OCCAM DR19 spectroscopic
  gradient is $-0.079 \pm 0.006$\,dex\,kpc$^{-1}$ \citep{Otto2026}.}
  \label{tab:gradient}
  \begin{tabular}{lccc}
    \hline
    $\eta_\mathrm{fit} \geq$ & $\nabla_{\!\feh}$ & $r$ & $N$ \\
    \hline
    \multicolumn{4}{l}{\textit{All clusters}} \\
    0.70 & $-0.051 \pm 0.002$ & $-0.51$ & 3789 \\
    0.80 & $-0.052 \pm 0.003$ & $-0.51$ & 3624 \\
    0.90 & $-0.052 \pm 0.003$ & $-0.50$ & 3146 \\
    0.95 & $-0.058 \pm 0.003$ & $-0.47$ & 2510 \\
    \multicolumn{4}{l}{\textit{XP-prior clusters}} \\
    0.70 & $-0.061 \pm 0.004$ & $-0.65$ & 765 \\
    0.80 & $-0.064 \pm 0.004$ & $-0.67$ & 716 \\
    0.90 & $-0.064 \pm 0.004$ & $-0.67$ & 600 \\
    0.95 & $-0.079 \pm 0.005$ & $-0.68$ & 439 \\
    \hline
  \end{tabular}
\end{table}

\begin{enumerate}

  \item \textit{Pipeline and priors.}  Each cluster is modelled in a
        four-dimensional parameter space $[\log(\mathrm{Age}/\mathrm{yr}),\,
        Z_\mathrm{ini},\, d_\mathrm{pc},\, E(G_\mathrm{BP}{-}G_\mathrm{RP})]$.
        Age priors are uniform over $\log(\mathrm{Age}/\mathrm{yr}) \in [6, 10]$
        for all clusters.  Distance priors are constructed as truncated Gaussians
        centred on the median Gaia DR3 parallax-inverted distance of cluster members,
        with a width reflecting the precision of that mean-distance estimate --- the
        observed member scatter reduced by the square root of the member count, since
        the prior constrains the cluster's single characteristic distance rather than
        the distances of individual stars. This provides an informative astrometric
        anchor while the isochrone likelihood constrains age, metallicity, and
        extinction; the reported distances are consequently astrometry-dominated,
        with a $\sim$2~per~cent uncertainty floor that neglects the spatially
        correlated \textit{Gaia} parallax systematic and should be treated as a
        lower bound in the outer disc (Section~\ref{sec:priors}).  Metallicity
        priors are informed by Gaia XP spectrophotometric \mbox{[Fe/H]} estimates
        for $1\,346$ clusters with at least three member stars, serving as an
        informative starting point rather than a hard constraint; a broad uniform
        prior $[-0.5, +0.5]$~dex is adopted otherwise.  Crucially, metallicity
        is treated as a fully free parameter throughout: the nested sampler explores the entire prior volume without anchoring to any external abundance-radius relation or spectroscopic metallicity catalogue; the metallicity priors (environment-conditioned uniform distributions or Gaia XP Gaussian, see Section~\ref{sec:priors}) are observationally motivated and deliberately broad, allowing the CMD likelihood to drive the posterior whenever the photometry provides sufficient discriminating power.  Extinction priors are anchored to SFD dust maps with a
        distance-dependent attenuation correction.  All posterior inferences were
        performed on the TRUBA high-performance computing facility using
        \textsc{dynesty}'s multi-ellipsoidal sampler with $N_\mathrm{live} = 400$
        live points.

  \item \textit{Sample and quality.}  Of the $5\,056$ processed clusters ($2\,156$
        AA-class and $2\,900$ AB-class), $3\,807$
        ($75.3$~per~cent) attain a high-quality fit flag $\eta_\mathrm{fit} \ge 0.80$,
        indicating well-constrained posteriors and reliable convergence.  The
        remaining clusters have broader posteriors, typically due to sparse
        membership, unfavourable photometric depth, or strong differential reddening;
        such entries are retained in the catalogue with their $\eta_\mathrm{fit}$
        flag, as broad posteriors reflect genuine photometric uncertainty rather
        than a failure of the inference and remain useful for statistical ensemble
        studies when appropriate quality thresholds are applied.

  \item \textit{Parameter distributions.}  The derived ages span
        $\log(\mathrm{Age}/\mathrm{yr}) = 6.60$--$9.77$ with a median of $8.35$
        (MAD $= 0.30$~dex), consistent with a disc population dominated by
        intermediate-age clusters.  Heliocentric distances range from $86$ to
        $17\,012$~pc, with a median of $2\,172$~pc, extending the well-sampled
        Solar neighbourhood out to the outer disc.  Photometric metallicities
        span $[\mathrm{Fe/H}] = -1.17$ to $+0.46$~dex (median $-0.001$~dex,
        MAD $= 0.061$~dex), and total extinctions $A_G$ reach up to
        $7.40$~mag (median $1.24$~mag).

  \item \textit{Validation against literature catalogues.} All residuals
        are defined as $\Delta X = X_\mathrm{ref} - X_\mathrm{this\,work}$.
        Agreement with \citet{CantatGaudin2020} ($N = 1\,230$ clusters in
        common) is close: $\Delta\log(\mathrm{Age}) = +0.022 \pm 0.259$
        (MAD $= 0.154$), $\Delta d = -44 \pm 330$~pc (MAD $= 57$~pc), and
        $\Delta A_G = -0.396 \pm 0.280$~mag (MAD $= 0.154$~mag).
        Comparison with \citet{Dias2021} ($N = 1\,045$) gives
        $\Delta\log(\mathrm{Age}) = +0.007 \pm 0.247$ (MAD $= 0.167$),
        $\Delta d = -224 \pm 339$~pc (MAD $= 123$~pc),
        $\Delta[\mathrm{Fe/H}] = +0.041 \pm 0.148$~dex (MAD $= 0.083$~dex),
        and $\Delta A_G = -0.160 \pm 0.264$~mag (MAD $= 0.138$~mag).
        Comparison with \citet{Hunt2023} ($N = 2\,337$) yields
        $\Delta\log(\mathrm{Age}) = -0.098 \pm 0.203$ (MAD $= 0.096$),
        $\Delta d = -179 \pm 285$~pc (MAD $= 101$~pc), and
        $\Delta A_G = -0.230 \pm 0.266$~mag (MAD $= 0.133$~mag),
        with no statistically significant systematic age bias.
        Comparison with \citet{Cavallo2024} ($N = 2\,308$) yields
        $\Delta\log(\mathrm{Age}) = -0.021 \pm 0.343$ (MAD $= 0.207$),
        $\Delta d = -274 \pm 579$~pc (MAD $= 153$~pc),
        $\Delta[\mathrm{Fe/H}] = +0.060 \pm 0.396$~dex (MAD $= 0.218$~dex),
        and $\Delta A_G = -0.108 \pm 0.329$~mag (MAD $= 0.173$~mag).
        Comparison with \citet{Bossini2019} ($N = 229$) yields
        $\Delta\log(\mathrm{Age}) = -0.051 \pm 0.242$ (MAD $= 0.147$),
        $\Delta d = -21 \pm 158$~pc (MAD $= 46$~pc), and
        $\Delta A_G = -0.125 \pm 0.213$~mag (MAD $= 0.135$~mag).
        Across all five photometric catalogues the systematic offsets in age
        remain below $0.10$~dex in absolute value, demonstrating the
        robustness of the PARSEC-based inference.

  \item \textit{Spectroscopic metallicity validation.}  Cross-matching with the
        OCCAM DR19 catalogue of spectroscopic open-cluster metallicities
        \citep{Otto2026} ($N = 132$ clusters in common) yields
        $\Delta[\mathrm{Fe/H}] = -0.021 \pm 0.132$~dex (MAD $= 0.054$~dex),
        confirming that the photometrically derived metallicities are consistent
        with independent high-resolution spectroscopic measurements to well within
        the systematic uncertainty of isochrone-based methods ($\lesssim 0.1$~dex).
        This agreement holds across a wide range of ages
        ($\log(\mathrm{Age/yr}) = 6.8$--$9.7$) and Galactocentric radii
        ($R_\mathrm{GC} = 5.7$--$14.0$~kpc), spanning the diversity of disc cluster
        environments represented in the present catalogue.

  \item \textit{Radial metallicity gradient.}  Using the catalogue's own
        metallicities we recover the disc radial [Fe/H] gradient
        (Section~\ref{sec:comparison:gradient}). Restricting the fit to the
        XP-informed clusters --- where the metallicity is genuinely constrained ---
        yields $\nabla_{\!\feh} = -0.064 \pm 0.004$\,dex\,kpc$^{-1}$ at
        $\eta_\mathrm{fit}\geq0.80$, consistent with recent open-cluster
        determinations ($-0.064\pm0.007$; \citealt{Spina2022}) and steepening
        towards the OCCAM DR19 spectroscopic value ($-0.079 \pm 0.006$\,dex\,kpc$^{-1}$;
        \citealt{Otto2026}) for the highest-quality clusters. Because the
        XP-informed metallicities inherit the \citet{Andrae2023} XP scale, this
        is a population-level consistency check that shows the pipeline preserves
        the spectroscopically-encoded gradient rather than an independent
        derivation; it confirms the catalogue's readiness for Galactic
        chemical-evolution studies, and shows that the gradient is diluted --- not
        by poor fits, but by the prior-dominated non-XP clusters --- underscoring
        that the informative metallicity signal resides in the XP subset.

  \item \textit{Scientific applications.}  The catalogue provides the largest
        homogeneous open-cluster parameter set to date derived with fully free
        photometric metallicity — $5\,056$ clusters processed with a single,
        uniform Bayesian pipeline on \textit{Gaia} DR3 data.  It constitutes
        a self-consistent reference baseline for a broad range of Galactic
        science: reconstruction of the age--metallicity and
        age--velocity-dispersion relations of the disc; chemical tagging and
        identification of co-natal stellar populations; mapping of the
        three-dimensional dust distribution; and calibration of stellar
        evolution models.  The photometric \mbox{[Fe/H]} values are derived
        without assuming any external abundance model, so they provide
        model-independent metallicity estimates that are directly comparable
        with spectroscopic surveys and can serve as photometric priors for
        forthcoming high-multiplex spectroscopic programmes (4MOST, WEAVE).
        The combination of precise \textit{Gaia} DR3 membership lists
        \citep{Perren2023} with Bayesian parameter estimation ensures
        that cluster-to-field contamination is minimised, making this catalogue
        particularly suited for statistical ensemble studies.

  \item \textit{Future prospects.}  The arrival of Gaia DR4, with its improved
        astrometric solutions, extended photometric calibrations, and
        radial-velocity measurements for a substantially larger fraction of
        cluster members, will enable further refinement of membership probabilities
        and tighter parallax-based distance priors.  Incorporation of ground-based
        and space-based spectroscopic surveys (e.g.\ 4MOST, WEAVE) will allow
        spectroscopic metallicity priors to be applied to a much larger fraction of
        the sample, breaking the photometric age--metallicity degeneracy for
        metal-poor and heavily reddened clusters.  Extension of the pipeline to
        the remaining UCC clusters of lower astrometric quality, as well as to
        embedded and very young clusters currently excluded by the membership
        quality cuts, represents a natural next step.

\end{enumerate}

The parameter catalogue is provided as a machine-readable table submitted with this article.
The complete set of \textsc{dynesty} posterior samples (HDF5 format,
one file per cluster) is archived on Zenodo
(\url{https://doi.org/10.5281/zenodo.23349618}), enabling full independent
reproduction of all results and supporting user-defined re-analyses of
individual cluster posteriors.  The pipeline source code, \textsc{ocnsfit}
(Open Cluster Nested Sampling Fit), is publicly available on GitHub
(\url{https://github.com/oplevne/ocnsfit}).
The tabulated posterior medians and credible intervals are designed to be
directly usable for disc structure and chemical-evolution studies, while
the published $\eta_\mathrm{fit}$ flags allow users to tailor quality
cuts to the specific requirements of their analysis.

\section*{Acknowledgements}
We thank the referee for a careful reading of the manuscript and for
constructive comments that improved the paper.
This work was supported by the Scientific and Technological Research Council of Turkey TUBITAK under grant 125F465 and the 2211-C (Furkan Akbaba) and 2219 (Olcay Plevne) programmes. This study was funded by the Scientific Research Projects Coordination Unit of the Istanbul University. Project number: FBA-2023-39380 and FDK-2025-41537. The numerical calculations reported in this paper were fully performed at TUBITAK ULAKBIM, High Performance and Grid Computing Center (TRUBA resources). This work has made use of data from the European Space Agency (ESA) mission \textit{Gaia}
(\url{https://www.cosmos.esa.int/gaia}), processed by the \textit{Gaia}
Data Processing and Analysis Consortium
(DPAC, \url{https://www.cosmos.esa.int/web/gaia/dpac/consortium}).
Funding for the DPAC has been provided by national institutions, in
particular the institutions participating in the \textit{Gaia}
Multilateral Agreement.
This research has made use of NASA's Astrophysics Data System
Bibliographic Services.
This work made use of the following software packages:
\textsc{Astropy} \citep{Astropy2013, Astropy2022},
\textsc{dynesty} \citep{Speagle2020},
\textsc{NumPy} \citep{Harris2020},
\textsc{SciPy} \citep{Virtanen2020},
\textsc{pandas} \citep{McKinney2010},
\textsc{scikit-learn} \citep{Pedregosa2011},
\textsc{Matplotlib} \citep{Hunter2007},
\textsc{corner} \citep{ForemanMackey2016},
\textsc{mwdust} \citep{Bovy2016} and \textsc{h5py}.

\section*{Data Availability}
The full parameter catalogue --- including posterior medians, 16th/84th
percentile credible intervals, the fit-quality indicator $\etafit$,
convergence flag, and log-evidence $\ln Z$ for all $5\,056$ clusters ---
is provided as a machine-readable table submitted with this article.
In addition to the posterior $G$-band extinction $A_G$, the published
table includes a derived visual-extinction column
$A_V = A_G/0.796$ (adopting $R_V = 3.1$ and the \citealt{Wang2019}
coefficients; see Section~\ref{sec:priors}), so that users can compare
directly against literature catalogues that report $A_V$ without having
to apply the conversion themselves.

The complete set of \dynesty{} posterior samples in HDF5 format
(one file per cluster, totalling $5\,056$ files) is archived
separately on Zenodo (\url{https://doi.org/10.5281/zenodo.23349618}).
These files contain the full nested-sampling chains and can be used to
reproduce all figures and statistics in this paper, or to carry out
independent analyses of individual cluster posteriors without re-running
the pipeline.

The pipeline source code, \textsc{ocnsfit}, is publicly available on GitHub
(\url{https://github.com/oplevne/ocnsfit}) under the MIT licence.

\bibliographystyle{mnras}
\bibliography{references}

\appendix

\section{Catalogue Column Descriptions}
\label{app:catalog}

Table~\ref{tab:catalog} lists all columns of the published open cluster
parameter catalogue together with their units and descriptions.
The catalogue contains $5\,056$ entries, one per cluster, and is made
available in full as a machine-readable table accompanying this article.
Each row provides posterior medians and asymmetric $1\sigma$ credible
intervals (16th and 84th percentiles) for the four fitted parameters
($\log\mathrm{Age}$, $[\mathrm{Fe/H}]$, $d$, $E(G_\mathrm{BP}{-}G_\mathrm{RP})$)
together with the derived $G$-band extinction $A_G$ and the
corresponding visual extinction $A_V = A_G/0.796$, the fit-quality
indicator $\etafit$, the nested-sampling convergence flag, and the
log-evidence $\ln Z$. For $\log\mathrm{Age}$ and age, the tabulated lower/upper
values are the 1$\sigma$ uncertainties (i.e.\ the offsets from the median to
the 16th/84th percentile); for all other parameters, the tabulated lower/upper
values are the 16th/84th percentiles of the posterior themselves, and the
corresponding uncertainty is obtained by subtracting the median.
\begin{table*}
  \centering
  \caption{Description of the published catalogue columns. The full table of
  $5\,056$ clusters is provided as a machine-readable supplementary file; the
  posterior HDF5 chains are archived on Zenodo. All uncertainties are the
  16th--50th and 50th--84th percentile intervals of the marginal posterior
  distributions.}
  \label{tab:catalog}
  \begin{tabular}{llp{11cm}}
    \hline
    Column & Unit & Description \\
    \hline
  \texttt{Cluster}          & ---       & Cluster identifier (semicolon-separated aliases; primary name is the first token) \\
  \texttt{source}           & ---       & UCC quality class: AA  or AB  \\
  \texttt{RA\_ICRS}         & deg       & Right ascension of the cluster centre (ICRS, \textit{Gaia} DR3) \\
  \texttt{DE\_ICRS}         & deg       & Declination of the cluster centre (ICRS, \textit{Gaia} DR3) \\
  \texttt{GLON}             & deg       & Galactic longitude $l$ of the cluster centre \\
  \texttt{GLAT}             & deg       & Galactic latitude $b$ of the cluster centre \\
  \texttt{R\_GC\_kpc}       & kpc       & Galactocentric radius, $R_\mathrm{GC}=(R_0^2+(d\cos b)^2-2R_0 d\cos b\cos l)^{1/2}$ with $R_0=8.2$\,kpc \citep{GravityCollab2019} \\
  \texttt{N\_members\_mask} & ---       & Number of member stars with membership probability $p \geq 0.75$ (used in the fit) \\
  \texttt{N\_members\_total}& ---       & Total number of member stars in the UCC membership file \\
  \texttt{logAge\_Med}      & dex       & Posterior median of $\log(\mathrm{Age/yr})$ \\
  \texttt{logAge\_ErrLo}    & dex       & Lower 1$\sigma$ uncertainty: median $-$ 16th percentile \\
  \texttt{logAge\_ErrHi}    & dex       & Upper 1$\sigma$ uncertainty: 84th percentile $-$ median \\
  \texttt{Age\_Med\_Gyr}    & Gyr       & Posterior median age in gigayears \\
  \texttt{Age\_ErrLo\_Gyr}  & Gyr       & Lower age uncertainty (Gyr) \\
  \texttt{Age\_ErrHi\_Gyr}  & Gyr       & Upper age uncertainty (Gyr) \\
  \texttt{FeH\_Med}         & dex       & Posterior median of $[\mathrm{Fe/H}]$ (photometric) \\
  \texttt{FeH\_Lo}          & dex       & 16th percentile of the $[\mathrm{Fe/H}]$ posterior \\
  \texttt{FeH\_Hi}          & dex       & 84th percentile of the $[\mathrm{Fe/H}]$ posterior \\
  \texttt{n\_XP}            & ---       & Number of member stars with a \textit{Gaia} XP spectrophotometric [Fe/H] estimate \\
  \texttt{feh\_prior}       & ---       & Metallicity prior type: \texttt{XP} (informative \textit{Gaia} XP Gaussian, $n_\mathrm{XP}\geq3$) or \texttt{uniform} (environment-dependent uniform); the [Fe/H] posterior is reliable only for \texttt{XP} clusters (Section~\ref{sec:results:properties}) \\
  \texttt{Z\_Med}           & ---       & Posterior median of initial metallicity $Z_\mathrm{ini}$ \\
  \texttt{Z\_Lo}            & ---       & 16th percentile of the $Z_\mathrm{ini}$ posterior \\
  \texttt{Z\_Hi}            & ---       & 84th percentile of the $Z_\mathrm{ini}$ posterior \\
  \texttt{Dist\_Med\_pc}    & pc        & Posterior median heliocentric distance \\
  \texttt{Dist\_Lo\_pc}     & pc        & 16th percentile of the heliocentric distance posterior \\
  \texttt{Dist\_Hi\_pc}     & pc        & 84th percentile of the heliocentric distance posterior \\
  \texttt{E\_BPRP\_Med}     & mag       & Posterior median colour excess $E(G_\mathrm{BP}{-}G_\mathrm{RP})$ \\
  \texttt{E\_BPRP\_Lo}      & mag       & 16th percentile of the $E(G_\mathrm{BP}{-}G_\mathrm{RP})$ posterior \\
  \texttt{E\_BPRP\_Hi}      & mag       & 84th percentile of the $E(G_\mathrm{BP}{-}G_\mathrm{RP})$ posterior \\
  \texttt{f\_inlier}        & ---       & Fit quality indicator $\etafit$ (fraction of CMD members consistent with the best-fit isochrone; see \S\ref{sec:quality}) \\
  \texttt{Converged}        & ---       & Nested-sampling convergence flag ($\Delta\ln Z < 0.01$) \\
  \texttt{LogZ}             & ---       & Nested-sampling log-evidence $\ln Z$ \\
  \texttt{LogZ\_err}        & ---       & Uncertainty of $\ln Z$ \\
  \texttt{A\_G\_Med}        & mag       & Posterior median $G$-band extinction $A_G = 1.890 \times E(G_\mathrm{BP}{-}G_\mathrm{RP})$ \citep{Wang2019} \\
  \texttt{A\_G\_ErrLo}      & mag       & 16th percentile of the $A_G$ posterior \\
  \texttt{A\_G\_ErrHi}      & mag       & 84th percentile of the $A_G$ posterior \\
  \texttt{A\_V\_Med}        & mag       & Derived visual extinction $A_V = A_G/0.796$ ($R_V = 3.1$; \citealt{Wang2019}), for direct comparison with literature catalogues reporting $A_V$ \\
  \texttt{A\_V\_ErrLo}      & mag       & 16th percentile of the $A_V$ posterior \\
  \texttt{A\_V\_ErrHi}      & mag       & 84th percentile of the $A_V$ posterior \\
    \hline
  \end{tabular}
\end{table*}

As an illustration of the tabulated quantities, Table~\ref{tab:catalog_excerpt}
lists the derived parameters for a representative selection of $30$ well-known
open clusters, ordered by age. This excerpt is provided for readability only;
the complete catalogue of all $5\,056$ clusters is available exclusively in
machine-readable form.

\begin{table*}
\centering
\caption{Derived parameters for a representative selection of $30$ well-known open clusters, ordered by age, as an illustrative excerpt of the full catalogue. The complete table of all $5\,056$ clusters --- including asymmetric $1\sigma$ credible intervals, $A_G$, $A_V$, the convergence flag, and $\ln Z$ --- is available in machine-readable form (CDS/VizieR and as a CSV in the data release). Uncertainties are the 16th/84th percentile credible intervals.}
\label{tab:catalog_excerpt}
\begin{tabular}{llccccc}
\hline
Cluster & Alias & $\log(\mathrm{Age/yr})$ & $[\mathrm{Fe/H}]$ & $d$ & $E(G_\mathrm{BP}{-}G_\mathrm{RP})$ & $\eta_\mathrm{fit}$ \\
        &       & (dex) & (dex) & (pc) & (mag) & \\
\hline
NGC\,6231 & -- & $6.75^{+0.47}_{-0.43}$ & $-0.12^{+0.29}_{-0.24}$ & $1649^{+33}_{-33}$ & $0.60^{+0.35}_{-0.39}$ & 0.93 \\
NGC\,2264 & -- & $6.76^{+0.49}_{-0.43}$ & $-0.01^{+0.34}_{-0.33}$ & $725^{+14}_{-14}$ & $0.54^{+0.45}_{-0.35}$ & 0.86 \\
Collinder\,69 & $\lambda$ Ori & $6.77^{+0.48}_{-0.43}$ & $+0.00^{+0.34}_{-0.34}$ & $400^{+8}_{-8}$ & $0.28^{+0.28}_{-0.19}$ & 0.93 \\
NGC\,6531 & M21 & $6.84^{+0.53}_{-0.46}$ & $-0.02^{+0.35}_{-0.33}$ & $1251^{+26}_{-24}$ & $0.40^{+0.33}_{-0.27}$ & 0.99 \\
NGC\,869 & h Per & $7.38^{+0.52}_{-0.58}$ & $+0.01^{+0.31}_{-0.35}$ & $2473^{+50}_{-49}$ & $0.73^{+0.21}_{-0.46}$ & 0.98 \\
NGC\,2547 & -- & $7.38^{+0.52}_{-0.51}$ & $-0.02^{+0.34}_{-0.33}$ & $387^{+8}_{-8}$ & $0.25^{+0.27}_{-0.17}$ & 0.95 \\
IC\,2391 & -- & $7.46^{+0.51}_{-0.50}$ & $-0.01^{+0.34}_{-0.34}$ & $151^{+3}_{-3}$ & $0.22^{+0.26}_{-0.15}$ & 0.90 \\
Trumpler\,10 & -- & $7.55^{+0.49}_{-0.51}$ & $-0.02^{+0.35}_{-0.32}$ & $435^{+9}_{-9}$ & $0.31^{+0.37}_{-0.22}$ & 0.99 \\
NGC\,1960 & M36 & $7.55^{+0.54}_{-0.54}$ & $-0.02^{+0.34}_{-0.33}$ & $1182^{+24}_{-23}$ & $0.42^{+0.36}_{-0.25}$ & 0.98 \\
NGC\,6405 & M6 & $7.61^{+0.52}_{-0.51}$ & $-0.01^{+0.34}_{-0.33}$ & $460^{+9}_{-9}$ & $0.33^{+0.34}_{-0.22}$ & 0.99 \\
NGC\,7654 & M52 & $7.85^{+0.56}_{-0.53}$ & $+0.03^{+0.32}_{-0.36}$ & $1643^{+33}_{-33}$ & $0.48^{+0.43}_{-0.34}$ & 0.72 \\
Melotte\,22 & Pleiades & $7.92^{+0.51}_{-0.50}$ & $-0.00^{+0.33}_{-0.33}$ & $136^{+3}_{-3}$ & $0.15^{+0.15}_{-0.10}$ & 0.89 \\
Blanco\,1 & -- & $8.06^{+0.53}_{-0.51}$ & $-0.03^{+0.36}_{-0.32}$ & $237^{+5}_{-5}$ & $0.14^{+0.15}_{-0.09}$ & 0.90 \\
NGC\,2422 & M47 & $8.09^{+0.52}_{-0.50}$ & $-0.02^{+0.34}_{-0.33}$ & $477^{+10}_{-10}$ & $0.31^{+0.35}_{-0.21}$ & 0.94 \\
NGC\,2516 & -- & $8.17^{+0.52}_{-0.51}$ & $+0.00^{+0.33}_{-0.33}$ & $412^{+8}_{-8}$ & $0.18^{+0.15}_{-0.12}$ & 0.94 \\
NGC\,1039 & M34 & $8.19^{+0.54}_{-0.53}$ & $-0.00^{+0.33}_{-0.33}$ & $500^{+10}_{-10}$ & $0.15^{+0.15}_{-0.10}$ & 0.95 \\
NGC\,6475 & M7 & $8.33^{+0.51}_{-0.51}$ & $-0.02^{+0.33}_{-0.32}$ & $279^{+6}_{-6}$ & $0.28^{+0.34}_{-0.19}$ & 0.98 \\
NGC\,2323 & M50 & $8.36^{+0.47}_{-0.55}$ & $-0.00^{+0.33}_{-0.33}$ & $995^{+20}_{-20}$ & $0.36^{+0.24}_{-0.21}$ & 1.00 \\
NGC\,2168 & M35 & $8.37^{+0.52}_{-0.54}$ & $-0.01^{+0.33}_{-0.33}$ & $869^{+17}_{-17}$ & $0.35^{+0.27}_{-0.22}$ & 0.99 \\
NGC\,3532 & -- & $8.46^{+0.55}_{-0.51}$ & $-0.10^{+0.28}_{-0.24}$ & $479^{+10}_{-9}$ & $0.28^{+0.36}_{-0.19}$ & 0.96 \\
NGC\,2632 & Praesepe & $8.65^{+0.56}_{-0.52}$ & $-0.00^{+0.33}_{-0.33}$ & $185^{+4}_{-4}$ & $0.14^{+0.14}_{-0.10}$ & 0.93 \\
NGC\,2437 & M46 & $8.68^{+0.37}_{-0.46}$ & $-0.03^{+0.26}_{-0.27}$ & $1675^{+34}_{-33}$ & $0.30^{+0.22}_{-0.16}$ & 0.99 \\
NGC\,6633 & -- & $8.80^{+0.48}_{-0.49}$ & $-0.11^{+0.27}_{-0.24}$ & $394^{+8}_{-8}$ & $0.27^{+0.22}_{-0.17}$ & 0.98 \\
NGC\,2099 & M37 & $8.86^{+0.39}_{-0.41}$ & $+0.03^{+0.26}_{-0.28}$ & $1485^{+30}_{-29}$ & $0.39^{+0.20}_{-0.17}$ & 1.00 \\
NGC\,2477 & -- & $9.06^{+0.42}_{-0.40}$ & $+0.02^{+0.27}_{-0.28}$ & $1449^{+28}_{-29}$ & $0.48^{+0.21}_{-0.20}$ & 0.99 \\
NGC\,752 & -- & $9.20^{+0.35}_{-0.40}$ & $-0.08^{+0.26}_{-0.22}$ & $441^{+9}_{-9}$ & $0.12^{+0.14}_{-0.08}$ & 0.98 \\
NGC\,2360 & -- & $9.22^{+0.36}_{-0.44}$ & $-0.11^{+0.26}_{-0.23}$ & $1090^{+21}_{-21}$ & $0.23^{+0.30}_{-0.15}$ & 0.99 \\
NGC\,7789 & -- & $9.32^{+0.35}_{-0.32}$ & $-0.01^{+0.27}_{-0.26}$ & $2089^{+41}_{-42}$ & $0.40^{+0.14}_{-0.15}$ & 0.99 \\
NGC\,2682 & M67 & $9.45^{+0.24}_{-0.38}$ & $-0.06^{+0.29}_{-0.26}$ & $867^{+17}_{-17}$ & $0.12^{+0.12}_{-0.08}$ & 0.98 \\
NGC\,188 & -- & $9.64^{+0.16}_{-0.28}$ & $+0.02^{+0.23}_{-0.24}$ & $1894^{+37}_{-38}$ & $0.16^{+0.10}_{-0.10}$ & 0.97
\\
\hline
\end{tabular}

\medskip
\raggedright
\textit{Note.} Representative excerpt only; the full machine-readable catalogue of $5\,056$ clusters accompanies this article (see Section~\ref{app:catalog} and the Data Availability statement).
\end{table*}

\section{Residuals of the literature comparison}
\label{app:resid}

\begin{figure*}
  \centering
  \includegraphics[width=0.95\textwidth]{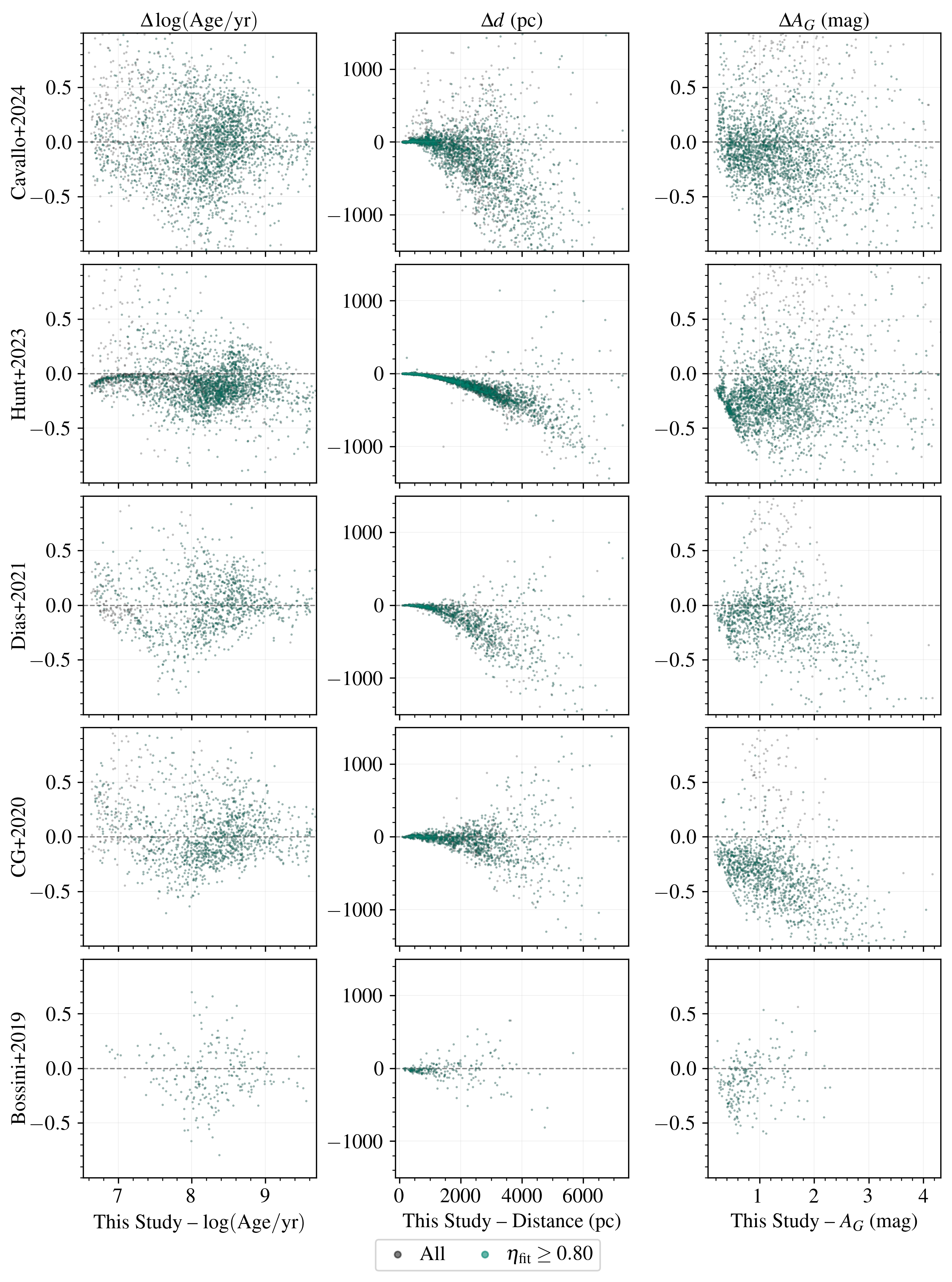}
  \caption{Residuals $\Delta X = X_\mathrm{ref} - X_\mathrm{this\,work}$ for
           age, distance, and $A_G$ as a function of the values derived in this
           work, for the same five photometric catalogues and in the same
           layout as Fig.~\ref{fig:compare_all4}.
           Grey points show the full sample;
           teal points show clusters with $\etafit \geq 0.80$.
           Dashed lines mark $\Delta X = 0$.}
  \label{fig:compare_resid}
\end{figure*}

Fig.~\ref{fig:compare_resid} shows the residuals of the comparison with the five
photometric reference catalogues discussed in Section~\ref{sec:comparison}, as a
function of the parameter values derived in this work. The corresponding offset
statistics are listed in Table~\ref{tab:comparison}.

\bsp	
\label{lastpage}
\end{document}